\documentclass[11pt]{article}

\usepackage[utf8x]{inputenc}
\usepackage[english]{babel}

\usepackage{amsmath}
\usepackage{graphicx}
\usepackage{titlesec}
\usepackage{float}
\usepackage{xcolor}
\usepackage{pdfpages}
\usepackage[colorlinks=true, allcolors=blue]{hyperref}
\usepackage{tikz}
\usepackage{indentfirst}
\usepackage{lipsum}  
\usepackage{csquotes}

\title{Dynamical Spreading Under Power Law Potential}
\author{
Ido Fanto
\thanks{School of Mathematical Sciences, Raymond and Beverly Sackler Faculty of Exact Sciences, Tel Aviv University, Tel Aviv 6997801, Israel.
Email: {\tt idofanto@mail.tau.ac.il}.}
}

\begin{document}
\begin{titlepage}
    \begin{center}
        \vspace*{0.5cm}
        \includegraphics{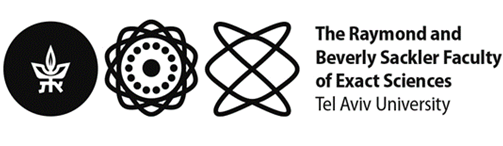}\\
        \vspace{0.3cm}
        {\Large SCHOOL OF PHYSICS AND ASTRONOMY}\\
        \vspace{1.2cm}
        {\huge Dynamical Spreading Under Power Law Potential\par}
        \vspace{0.8cm}
        {\Large Thesis submitted in partial fulfillment of the requirements for the M.Sc. degree in the School of Physics and Astronmy, Tel Aviv University\par}
        \vspace{0.8cm}
        {\Large By}\\
        \vspace{0.3cm}
        {\LARGE \textbf{Ido Fanto}}\\
        \vspace{1.2cm}
        {\Large The research work for the thesis has been carried out}\\
        \vspace{0.3cm}
        {\Large under the supervision of}\\
        \vspace{0.3cm}
        {\huge Dr. Naomi Oppenheimer}
    \end{center}
\end{titlepage}

\newpage
\begin{abstract}

In this thesis, we examine the dynamic spreading of a dense overdamped suspension of particles under power law repulsive potentials,  often called Riesz gases. That is, potentials that decay with distance as $1/r^k$ where $k\in (-2,\infty]$. Depending on the value of $k$ relative to the system's spatial dimension $D$, the potentials are categorized as short-ranged for $k > D$, and long-ranged when $k \leq D$. Such systems naturally occur in contexts involving particle suspensions, granular media, and charged systems, where interactions can be influenced by physical fields that decrease over distance.\\
Our analytical findings reveal that the particles spread in a self-similar form, with the radius growing with time as $t^\frac{1}{k+2}$. The theoretical predictions derived for a general dimension $D$, are verified by numerical simulations involving thousands of particles in free space, in both one and two dimensions. Furthermore, the simulations not only confirm our analytical results but also reveal a rich diversity of behaviors depending on the value of $k$. We demonstrate that the density profiles differ significantly depending on whether $k$ is larger than, smaller than, or equal to $D-2$, where $D$ is the dimension. For $k>D-2$ the density is centered in the middle and we also notice a Wigner lattice emerging as a result of the repulsive interactions, for $k = D-2$, density is uniform and for $k<D-2$, density is centered at the edges. This new classification indicates that the long/short-range classification is insufficient for predicting the density profile of the suspension. When $k<D-2$, we observed an interesting phenomenon when two or more suspensions are placed near each other: a particle-free zone is formed where the two populations meet, resembling structures of bubbles.\\
By bridging analytical theory, numerical simulations, and preliminary experimental observations, this thesis provides a comprehensive understanding of the spreading dynamics in overdamped Riesz gases. Our findings enrich the fundamental knowledge of particle systems with power-law interactions, and also offer insights into a wide range of applications, from material science to biological systems, where such interactions play a major role.
\end{abstract}

\newpage
\tableofcontents
\newpage

\section{Introduction}
\label{intro}

Power law potentials are an important part of our lives; It is the reason why the earth orbits the sun, it is how charged particles interact with each other, it even governs the interaction between molecules; as such, it is no wonder that it has been the subject of many studies. In this thesis, we will analyze the behavior of an overdamped suspension of particles interacting with each other under various power law potentials. We will begin by providing an introduction to the theoretical knowledge used in this work (Sec.~\ref{intro}), covering topics such as dimensional analysis, self-similarity, nonlinear diffusion, and Riesz gas. Afterward in Sec.~\ref{results}, we will show simulations of a suspension containing thousands of particles interacting with various power-law potentials. We will also present the resulting density profiles of the suspensions and examine how the radius of the suspension grows with time. We will show from the simulations that these systems behave in a self-similar manner. We will then support our analysis with analytical calculations. Additionally, we will illustrate how three vastly different density profiles emerge based on different scaling of the power law potential between particles. Lastly, we present simulations of collisions between two or more suspensions and analyze the significantly different results of the collision due to the power law potential. We will end by outlining future possible work including preliminary results from experiments of paramagnetic particles under an external magnetic field. 

\subsection{Dimensional analysis and Buckingham $\pi$ theorem}
Dimensional analysis is one of the most important tools in a scientist's toolbox, and among the first checks a scientist performs when deriving a new equation. In its very basic form, the analysis involves ensuring that each of the terms in the equation has the same dimension in order to ensure the equation makes physical sense. Aside from checking the dimensionality of derived equations, it is important to ensure that our physical laws are universally correct regardless of the unit system being used. The analysis can also give insight and scaling of important qunatities. The Buckingham $\pi$ theorem is a key method in this regard, providing a way to systematically reduce the number of unknowns in an equation. The theorem states that if a physically meaningful equation with 'n' physical variables and 'k' physical dimensions --- which are the basic units that are required to span the space of solutions. In mechanical problems, this number is usually $k=3$ --- length ($L$), mass ($M$), and time ($T$). In MKS --- meter, kilogram, and second, as the name implies. The equation can then be transformed into a dimensionless equation with 'p' dimensionless variables, where 'p' is given by
\begin{align}
    p=n-k.
\end{align}
A simple example to illustrate the usefulness of The Buckingham $\pi$ theorem follows. Say we are interested in finding the period $T$ of a simple pendulum. The mass of the pendulum is $m$, and its length is $l$, the pendulum's Equation of Motion (EOM) can be written using Newton's second law
\begin{align}
\label{eqPendulum}
    ml\Ddot{\theta}=-mg\sin{\theta}.
\end{align}
This equation has five physical variables --- $m, l, \theta, g, t$. There are three physical dimensions ($T, L, M$). According to Buckingham $\pi$ theorem, these five variables can be transformed to two dimensionless variables. Since the period has units of time and only $g$ depends on time, the period of the pendulum must depend on $g$. In order to get units of time from $g$ we must divide a length unit by $g$. That is  $T=\sqrt{\frac{l}{g}}f(\theta_{0})$ where $f(\theta_{0})$ can be a function or constant of another dimensionless quantity $\theta_{0}$. Solving Eq.~\ref{eqPendulum} for small angles gives the exact expression $T=2\pi\sqrt{\frac{l}{g}}$. To find the value of $2\pi$, we could have conducted an experiment where we measured the period of the pendulum for different initial angles, plotting the data, and obtaining the relevant coefficient from the slope. This demonstrates the power of using dimensional analysis as we have found the general form of the period of a pendulum without needing to solve the EOM.

\subsection{Self Similarity}

Self-similarity can be thought of as a form of symmetry that a system possesses. Just like we say that a circle is symmetric to operations such as rotation and reflection since the circle remains unchanged, an object is said to be self-similar if, upon re-scaling with respect to some variable or combination of variables (e.g. spatial rescaling is equivalent to zooming in or out), the overall shape of the object is maintained. Examples of self-similarity are fractals, both mathematical fractals such as the Mandelbrot set\cite{branner1989mandelbrot}, and in real-life fractals like coastlines\cite{Kappraff1986THEGO} or blood vessel networks.
\begin{figure}[H]
\centering
\includegraphics[width=0.3\linewidth]{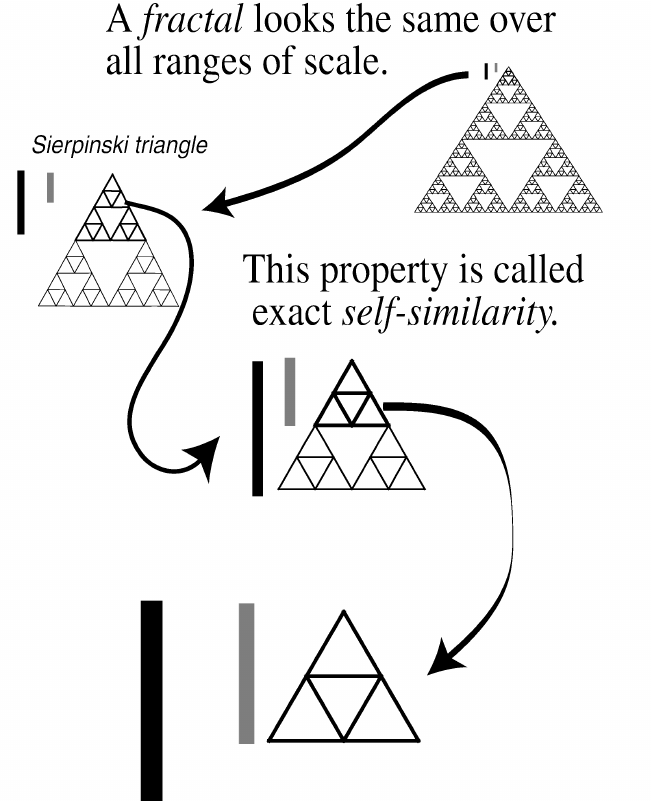}
\caption{\label{fig:A-depiction-of-geometric-exactly-self-similar-fractal-objects-The-Sierpinski-triangle.png}An example of self similarity in the Sierpinski triangle \cite{serpinski}}
\end{figure}
Self similarity can also occur in a dynamical process if an observed quantity, $n(r,t)$, exhibits dynamical scaling. To check if $n(r,t)$ has dynamical scaling, it is sufficient to show that it can be rewritten in the following way
\begin{equation}
\ n(r,t)\sim t^{\gamma }f\left(\frac{r}{t^{\beta }}\right),
\end{equation}
where $r$ and $t$ are space and time coordinates, $\gamma$ and $\beta$ are the scaling exponents, and $f$ is the scaling function. Under such re-scaling, values of $n$ for different times will collapse on the same curve.
This concept of self-similarity in dynamical systems is crucial for understanding phenomena such as phase transitions \cite{Stinchcombe1989FractalsPT}, critical phenomena in statistical mechanics \cite{Evans1994ObservationOC}, and certain turbulent flows in fluid dynamics \cite{Klewicki2013SelfsimilarMD}.

\subsection{Continuity Equation}
The continuity equation is a fundamental principle in various fields of physics. There are several methods to prove the continuity equation. Here we will undertake a relatively simple and basic approach. We will begin by examining an arbitrary volume within a fluid, where the volume's shape and size remain constant, but the surrounding fluid can freely enter and exit the volume, potentially changing the fluid's density within the volume over time. This description can be expressed mathematically as follows
\begin{equation}
\int \frac{\partial \rho }{\partial t}dV=-\int \rho \textbf{v}\cdot \textbf{n}dS,
\end{equation}
where \textbf{v} is the velocity of the fluid and \textbf{n} is the vector normal of the surface area of the volume and $dS$ is the surface integral on the shape. the divergence Theorem can now be used on the right hand side and rewrite it as
\begin{equation}
\int (\frac{\partial \rho }{\partial t}+\nabla \cdot(\rho \textbf{v}))dV=0.
\end{equation}
Since this equation needs to be satisfied for any arbitrarily chosen volume  we conclude that  the integrand must vanish and we are left with
\begin{equation}
\label{Continuity Equation}
 \frac{\partial \rho }{\partial t}+\nabla \cdot(\rho \textbf{v})=0.
\end{equation}
Which is known as the continuity equation.
\begin{figure}[H]
\centering
\includegraphics[width=0.5\linewidth]{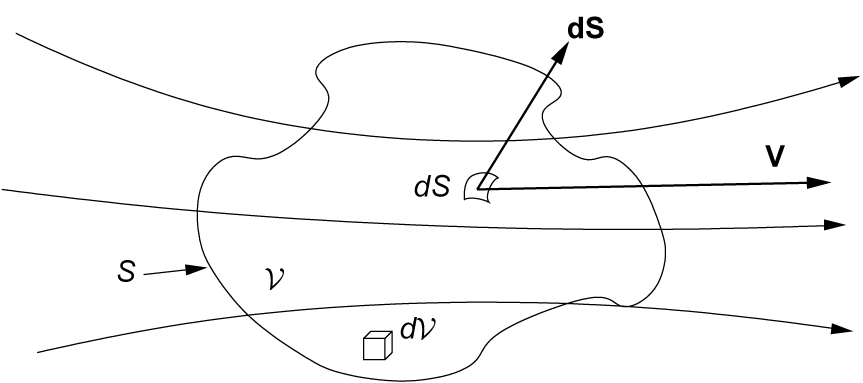}
\caption{\label{fig:con eq ills}An illustration of flux entering and exiting an arbitrarily chosen volume \cite{anderson2007fundamentals}}
\end{figure}

\subsection{Diffusion Equation}
 Diffusion was first observed in 1827 by Robert Brown. Brown noticed that pollen particles in water, when viewed through a microscope, move around randomly. At that time, the existence of molecules was not fully understood, so it was unknown what caused the pollen to move in this life-like manner. It wasn't until 1905 that Albert Einstein published a paper on Brownian motion, in which he suggested that the pollen particles moved as a result of water molecules hitting them from all directions. The direction in which the pollen particle moved was determined by the direction that absorbed the most hits from the water molecules, leading to the seemingly random motion of the pollen.
We encounter diffusion in everyday life, such as the way dye spreads in a still liquid or how heat is conducted throughout a solid. 

We will now derive Fick's diffusion equation from the continuity equation, Eq.~\ref{Continuity Equation}. We assume an overdamped system of an ideal gas, where the velocity is proportional to the force through a constant mobility $\mathbf{v}=\mu \mathbf{F}$. The force, in this case, is the gradient of the potential $U$, which for an ideal gas is $U=k_{B}T\ln{\rho}$, where $\rho$ is the density. By substituting these expressions into Eq.~\ref{Continuity Equation}, we obtain
\begin{align}
\label{Diffusion Equation}
    \frac{\partial \rho }{\partial t}+\vec{\nabla} \cdot(\rho \mathbf{v})= \frac{\partial \rho }{\partial t}+\vec{\nabla} \cdot(\rho \mu &\mathbf{F})= \frac{\partial \rho }{\partial t}-\vec{\nabla} \cdot(\rho \mu \vec{\nabla} k_{B}T\ln{\rho})=\frac{\partial \rho }{\partial t}-\vec{\nabla} \cdot( \mathcal{D} \vec{\nabla} \rho)=0\\
    & \frac{\partial \rho }{\partial t} =\vec{\nabla} \cdot( \mathcal{D} \vec{\nabla} \rho) \notag.
\end{align}
Eq.~\ref{Diffusion Equation} is called the diffusion equation and $\mathcal{D}=\mu  k_{B}T$ is the diffusion coefficient.

There are two main ways to analyze diffusion: the macroscopic picture and the microscopic picture. Both approaches yield the same results, but the formalism differs. We will review both approaches here, as later we will perform a similar procedure for repulsive particles and switch from the microscopic to the macroscopic picture. Let us begin with the macroscopic viewpoint. To do so, we first write the diffusion equation (Eq.~\ref{Diffusion Equation}) in 1D.
\begin{align}
    \frac{\partial P(x,t)}{\partial t} =\mathcal{D}\frac{\partial^2 P(x,t)}{\partial x^2},
\end{align}
where $P(x,t)$ is the probability of finding a particle at position $x$ at time $t$. the density $\rho(x,t)$ can be defined as $\rho(x,t) = NP(x,t)$. Since the probability of finding a particle is time-dependent, we can assume that its moments are also time-dependent. For example, the first moment is given by $\left< x(t) \right> = \int_{-\infty}^{\infty} xP(x,t)dx$. Additionally, we know that the particle exists somewhere in space, so we have $\int_{-\infty}^{\infty} P(x,t)dx = 1$.

Assuming the particle is initially located at the origin at $t=0$, the first moment must be zero because the forces acting on the particle are random, which implies that over time, these forces should cancel each other out. The second moment, however, is not zero
\begin{align}
    &\frac{\partial \left< x^{2}\right>}{\partial t}=\mathcal{D}\int_{-\infty}^{\infty}x^{2}\frac{\partial^2P(x,t) }{\partial x^2}dx = \mathcal{D}x^{2}\frac{\partial P(x,t)}{\partial x}\Big|_{-\infty}^\infty - \mathcal{D}\int_{-\infty}^{\infty}2x\frac{\partial P(x,t)}{\partial x}dx \nonumber \\
   & = -\mathcal{D}2xP(x,t)\Big|_{-\infty}^\infty  + 2\mathcal{D}\int_{-\infty}^{\infty} P(x,t)dx \label{eqSecondMoment} \\
   & \left< x^{2}\right> = 2\mathcal{D}t. \notag
\end{align}
We used integration by parts twice in Eq.~\ref{eqSecondMoment} and assumed that the probability and its derivative are zero at $\pm\infty$. We can see that the characteristic length scale for diffusive processes grows as $\sqrt{t}$, unlike ballistic motion which is proportional to $t$.\\

We will now look at the microscopic picture of diffusion. 
Let us start by writing the forces acting on a single Brownian particle, i.e. Newton's 2nd law,
\begin{equation}
\label{langevin Eq}
\dot{v}=-\frac{\gamma }{m}v +\frac{\sqrt{c}}{m}R(t).
\end{equation}
Equation~\ref{langevin Eq} is known as Langevin equation, where $\gamma$ is the friction coefficient of the liquid and $R(t)$ represents the random collisions of molecules with the Brownian particle. Were there no collisions, the velocity of the Brownian particle would decrease to zero on a time scale of $t \sim \frac{m}{\gamma}$. Since the collisions are random, the average of the force acting on a Brownian particle vanishes. We will assume that its time correlation is given by
\begin{equation}
\left<R(t)R(t')\right>=\delta(t-t').
\end{equation}
We assume that $R(t)$ is completely defined by its average and its correlation function. The parameter $c$ in Eq.~\ref{langevin Eq} is not known a priori, however we can find it using the equipartition theorem on the kinetic energy of the degrees of freedom of a liquid in equilibrium. At long times, $t\gg\frac{m}{\gamma}$, this gives $\left<v^{2}\right>=\frac{k_{B}T}{m}$. From Eq.~\ref{langevin Eq}, the velocity and its second moment are then given by,
\begin{align}
&v(t)=v_{0}e^{-\frac{\gamma }{m}t} + \frac{\sqrt{c}}{m}\int dt'e^{-\frac{\gamma }{m}(t-t')}R(t') = \frac{\sqrt{c}}{m}\int dt'e^{-\frac{\gamma }{m}(t-t')}R(t') \label{eqV2} \\
& \left<v^{2}(t) \right> = \frac{c}{m^{2}}\int_{0}^{t} dt' \int_{0}^{t} dt''e^{-\frac{\gamma }{m}(t-t')}e^{-\frac{\gamma }{m}(t-t'')}\delta (t'-t'')= \frac{c}{m^{2}}\int_{0}^{t} dt' e^{-\frac{2\gamma }{m}(t-t')} \notag \\
&  \Rightarrow \frac{k_{B}T}{m} = \frac{c}{2m\gamma}  \notag &\\
& \Rightarrow c= 2k_{B}T\gamma\notag .
\end{align}

Equation~\ref{eqV2} shows that the random forces that act on the Brownian particle are directly related to the friction as a result of the kinetic energy. At long time scales, the velocity reaches a steady state
\begin{align}
    v(t) = \frac{\sqrt{c}}{\gamma}R(t) = \sqrt{\frac{2k_{B}T}{\gamma}}R(t)=\sqrt{2\mathcal{D}}R(t),
\end{align}
where $\mathcal{D}=\frac{k_{B}T}{\gamma}$ is the diffusion coefficient as before since $\gamma = 1/\mu$. Solving for $\left<x^{2}(t) \right>$ instead, yields the mean squared displacement of the particle,
\begin{equation}
    \left<x^{2}(t) \right>=2\mathcal{D}t.
\end{equation}
Note that the same result was derived in the macroscopic picture, as expected.

The number of particles $N$ is conserved in time. This can be written as an integration over the density, $\rho(r,t)$, over space
\begin{equation}
\label{Constraint_N}
\ N=\int_{0}^{\infty}\rho(r,t)S_{D-1}r^{D-1}dr,
\end{equation}
where $S_{D-1}$ is the surface area of a $D-1$ dimensional unit sphere.\\
The solution of the diffusion equation, Eq.~\ref{Diffusion Equation}, with the constraint given by Eq.~\ref{Constraint_N} and an initial concentration of a delta function at the origin is well known to be a gaussian
\begin{align}
\label{Linear_diff_sol}
    \rho(r,t) = \frac{N}{(4\pi \mathcal{D}t)^{\frac{D}{2}}}\exp (-\frac{r^{2}}{4\mathcal{D}t}),
\end{align}
where $D$ is the dimension, and $\mathcal{D}$ is the diffusion coefficient. Let us note that this solution has a self-similar form of
\begin{align}
\label{self similar ansatz}
    \rho(r,t)=At^{\gamma}f\left(\frac{Br}{t^{\beta}}\right).
\end{align}
Could we have guessed the general form of the solution before solving Eq.~\ref{Diffusion Equation}? The answer is yes. For that goal, we will follow closely the framework used by Leal in~\cite{Leal_2007}. Starting from the diffusion equation, Eq.~\ref{Diffusion Equation}, and using dimensional analysis. All terms have the same dimensionality, thus 
\begin{align}
    \left[\frac{\rho}{t}\right]\sim\left[\frac{\mathcal{D}\rho}{r^{2}}\right]\Rightarrow \left[\frac{r^{2}}{\mathcal{D}t}\right]\sim[1].
\end{align}
From the above scaling, we can notice that the solution needs to be of the form of a dimensionless variable $\eta=\frac{Br}{t^{\beta}}$. However, we must also satisfy the constraint on the number of particles given by Eq.~\ref{Constraint_N}. Since the shape of the concentration depends on $\eta=\frac{Br}{t^{\beta}}$, the only way to satisfy the constraint is if the amplitude of the concentration is proportional to $t$ to some power, leading us back to the same form as Eq.~\ref{self similar ansatz}.

Both the diffusion equation, Eq.~\ref{Diffusion Equation}, and the constraint, Eq.~\ref{Constraint_N}, can be transformed from $(r,t)$ to $\eta$ by plugging the self-similar ansatz given by Eq.~\ref{self similar ansatz} in Eqs.~\ref{Diffusion Equation} and~\ref{Constraint_N}. Doing so will give the scaling exponents $\gamma$ and $\beta$. If the resulting governing equation after transforming to $\eta$ dependence is linear, we can expect to get an analytical closed form of $f(\eta)$; however, that will not always be the case.

Let us start by finding $\gamma$. To do so, we plug the self-similar form, Eq.~\ref{self similar ansatz}, into the constraint on the number of particles, Eq.~\ref{Constraint_N}, and switch variables from $(r,t)$ to $\eta$
\begin{align}
    N = S_{D-1}\int \rho r^{D-1}dr = S_{D-1}t^{\gamma+\beta D} \int f(\eta) \eta ^{D-1} d\eta.
\end{align}
Since the number of particles does not depend on time, we must demand that the exponent of $t$ is zero,
\begin{align}
\label{finding gamma}
    \gamma = -\beta D.
\end{align}
We then plug Eq.~\ref{finding gamma} and Eq.~\ref{self similar ansatz} into Eq.~\ref{Diffusion Equation} to get
\begin{align}
\label{governing eq for f in diff}
      -\beta \left(D\eta^{D-1} f(\eta) + \eta^{D} \frac{d f}{d \eta}\right)=\mathcal{D}B^{2}t^{1-2\beta}\frac{d }{d \eta}\left(\eta^{D-1}\frac{d f}{d \eta}\right),
\end{align}
where $\mathcal{D}$ is the diffusion coefficient and $D$ is the dimension. Since we assumed that $f$ is a function that only depends on $\eta$, there cannot be explicit time dependence in Eq.~\ref{governing eq for f in diff}. This lets us infer both $B$ and $\beta$
\begin{align}
\label{beta and b for diffusion}
    &\beta =\frac{1}{2}\\
    & B^{2}=\frac{1}{2\mathcal{D}}. \notag
\end{align}
The equation for $f(\eta)$ reads
\begin{align}
    \frac{d}{d \eta}\left(\eta^{D}f(\eta)\right) +\frac{d }{d \eta}\left(\eta^{D-1}\frac{d f}{d \eta}\right)=0.
\end{align}
The general solution to this equation regardless of the dimension, $D$ is
\begin{align}
\label{f for diffusion}
    f(\eta) = \exp\left(-\frac{\eta^{2}}{2}\right).
\end{align}
Plugging Eq.~\ref{f for diffusion} into Eq.~\ref{Constraint_N} and finding $A$
\begin{align}
\label{A for diffusion}
    A=\frac{N}{(4\pi \mathcal{D})^{\frac{D}{2}}}.
\end{align}
Plugging Eqs.~\ref{finding gamma}, \ref{beta and b for diffusion},\ref{f for diffusion} and \ref{A for diffusion}   back into Eq.~\ref{self similar ansatz} we indeed get the known solution of a delta function at the origin spreading diffusivity, Eq.~\ref{Linear_diff_sol}.

\subsection{Non Linear Diffusion} \label{Non Linear Diffusion}
Non-linear diffusion is utilized in various scientific fields. For example, diffusive particles that also interact by some potential answer to a non-linear diffusion equation \cite{PhysRevE.98.032138}. In image processing, it is employed to effectively "de-noise" an image \cite{Weickert1997ARO}. In fluid dynamics, it is used to simulate the flow of a fluid in a porous media \cite{Vzquez2006SmoothingAD}, and in biology, it is applied to model population dynamics \cite{MacCamy1981APM}. What connects all of the above examples is that the diffusion coefficient depends on the concentration. We will again be following the framework used by Leal in~\cite{Leal_2007}. We will start by revisiting Eq.~\ref{Diffusion Equation}, this time using a diffusion coefficient that is a power-law of the density, 
\begin{align}
    \mathcal{D}(\rho)=\mathcal{D}_{0}\left(\frac{\rho}{\rho_{0}}\right)^{n}.
\end{align}
We get the following nonlinear equation in $D$ dimensions
\begin{align}
\label{non linear diff}
    \frac{\partial \rho}{\partial t}=\frac{\mathcal{D}_{0}}{\rho_{0}^{n}r^{D-1}}\frac{\partial }{\partial r}\left(\rho ^{n}r^{D-1}\frac{\partial \rho}{\partial r}\right).
\end{align}
We now plug the self-similar ansatz, Eq.~\ref{self similar ansatz} in Eq.~\ref{non linear diff}. The constraint on the number of particles remains the same, so we will also use the value of $\gamma=-\beta D$ from Eq.~\ref{finding gamma}.  The LHS of Eq.~\ref{non linear diff} in terms of $\eta$ is
\begin{align}
    \frac{\partial \rho}{\partial t} = -A\beta t^{\beta D -1}\left( D f(\eta) + \eta \frac{d f}{d \eta}\right),
\end{align}
The RHS of Eq.~\ref{non linear diff} in terms of $\eta$ is
\begin{align}
    \frac{\mathcal{D}_{0}}{\rho_{0}^{n}r^{D-1}}\frac{\partial }{\partial r}\left(\rho ^{n}r^{D-1}\frac{\partial \rho}{\partial r}\right)=\frac{\mathcal{D}_{0}B^{2}A^{n+1}}{\rho_{0}^{n}}t^{-\beta Dn -2\beta -\beta D }\left(\eta^{1-D}\frac{d }{d \eta}\left(\eta^{D-1}f^{n}(\eta)\frac{d f}{d \eta}\right)\right).
\end{align}
Again  we remember that the governing equation for $f(\eta)$ can only depend on constants or $\eta$, we find that
\begin{align}
\label{beta for non linear diff}
    \beta = \frac{1}{2+ nD},
\end{align}
$B^{2}$ is then defined as
\begin{align}
    B^{2}=\frac{\beta n \rho_{0}^{n}}{2 \mathcal{D}_{0}A^{n}}.
\end{align}
We get the following governing equation for $f$
\begin{align}
\label{governing eq for f in non linear diff}
     \frac{d}{d \eta}\left(\eta^{D}f(\eta)\right)  + \frac{n}{2}\left[\frac{ }{d \eta}\left(\eta^{D-1}f^{n}(\eta)\frac{ f}{ \eta}\right)\right]=0.
\end{align}
The solution to Eq.~~\ref{governing eq for f in non linear diff} is
\begin{align}
\label{f for non linear diff}
    f(\eta)=(1 - \eta^{2})^{\frac{1}{n}}.
\end{align}
The density given by Eq.~\ref{f for non linear diff} is vastly different from the solution to the linear diffusion, Eq.~\ref{f for diffusion}. Most notably, perhaps, is that the density cannot be negative, $f(\eta)>0$, so that the solution above only makes physical sense for $\eta<1$. The mathematical term for this solution is "compact support" --- the solution is strictly zero beyond some cutoff. This is in contrast to the case of linear diffusion where the solution, Eq.~\ref{f for diffusion}, has the form of an exponent, which has infinite tails.

\subsection{Riesz Gas} \label{Riesz Gas}
A Riesz gas \cite{lewin2022coulomb} refers to a system of particles that interacts with each other by a power law potential $U(r)\propto \frac{1}{r^k}$, the exponent $k$ determines the nature of the interaction. Riesz gas systems are abundant, for example the case of $k=D-2$ describes a coulomb potential and a gravitational potential in $D$ dimensions.
Since Riesz gas systems are so prevalent, different research fields have come up with their own definition of what a Riesz gas is and how to study it. There are, however, two main parameters that determine the behavior of a Riesz gas --- the first being $k$ which we mentioned above and usually takes values between $(-2,\infty]$, the other one being $\xi=\frac{\rho^{\frac{k}{D}}}{T} \in [0,\infty]$ where $\rho$ is the density of particles in space, $\xi$ is often referred to as the intensity of the system, the temperature $T$ dependence in $\xi$ is there to give the system an element of randomness.

The potential is defined in such a way that the particles interact with each other through a repulsive potential.
\begin{align}   
    U(r) = \begin{cases} \frac{1}{r^k} &\mbox{for } k > 0 \\
    -\log(r) & \mbox{for } k=0 \\
    -\frac{1}{r^k} & \mbox{for } -2<k<0 .
    \end{cases}
\end{align}
It is possible to classify the interaction of particles by a Riesz potential by the range of the potential --- for $k>D$ the interaction is short-ranged, usually this means that particles interact only with their nearest neighbors; for $k<D$ the interaction is long-ranged, which means that every particle interacts with every other particle. These definitions arise as a result of whether the integration over all space of the potential $U(r)\propto \frac{1}{r^k}$ diverges (which leads to the classification of long-range interaction), or whether the integration converges (leading to short-range interaction).

\section{Results}
\label{results}
\begin{figure}[H]
\centering
\includegraphics[width=1\linewidth]{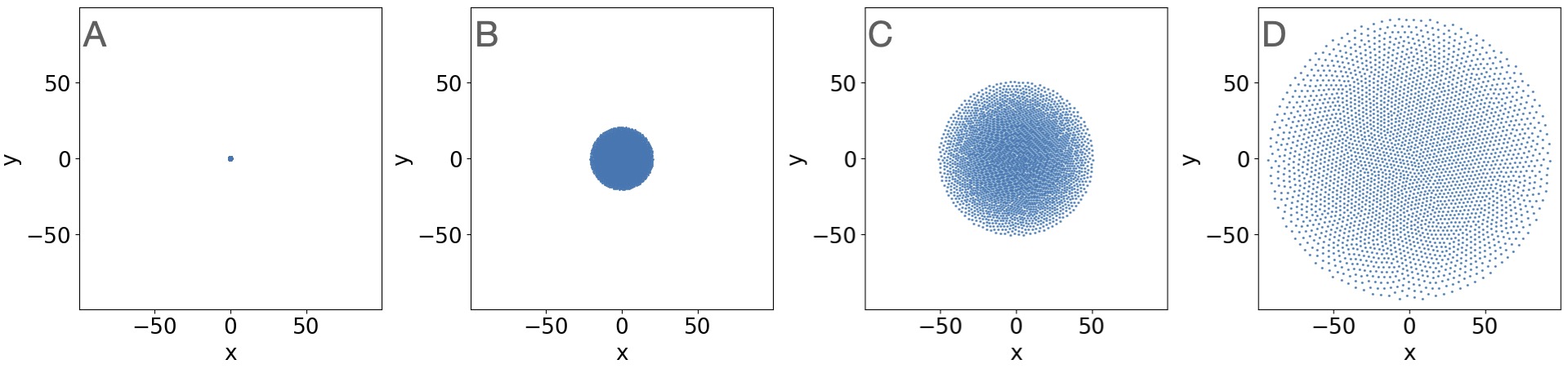}
\caption{\label{fig:inti} (A) Initial configuration of the suspension. Configuration of the suspension after (B) 500 time steps, (C),( 750 time steps.}
\end{figure}

In this thesis, we will study the behavior of particles in free space when subject to a repulsive power law potential of the form $U(r)=\frac{1}{r^k}$. We will start by performing molecular dynamics simulations of a dense suspension of repulsive particles. We will then try to formulate coarse-grained differential equations and derive analytically the dynamics of the suspension. In simulations, we focus on one and two dimensions, though the analytic results apply to any dimension. We study systems in the overdamped regime, which means that the velocity is directly proportional to the force through constant mobility $\mu$, $\textbf{v} = \mu {\textbf{F}} \propto\frac{1}{r^{k+1}}$. The velocity of each particle is determined by the contribution to the force acting on it from all other particles. For a system of $N$ particles, the velocity of the $i^{\rm th}$ particle can be written in the following discrete form
\begin{align}
\label{discrete velocity}
    \mathbf{v_{i}(r_{i})}=\mu\mathit{\sum_{j\neq i}^{N}}\frac{1}{\left|\mathbf{r_{i}-r_{j}} \right|^{k+1}}\mathbf{\hat{r}_{ij}}.
\end{align}
We begin by solving Eq.~\ref{discrete velocity} numerically using an 8th-order Runge-Kutta method with an adaptive timestep. We place thousands of point-like particles in a circular arrangement with $R=1$ in a uniform configuration in free space. We then calculate the velocity of each particle of the suspension by propagating Eq.~\ref{discrete velocity} in time. The optimal time step is chosen by the code and can change between steps. We repeat this process until we reach a predetermined time $t_{end}$.
When the simulation is complete, we have the positions of each particle at every time step.
Figure~\ref{fig:inti} gives four snapshots from a simulation of 3000 particles with $k=2$ --- the snapshots show the spreading of the suspension.

\begin{figure}[H]
\centering
\includegraphics[width=0.6\linewidth]{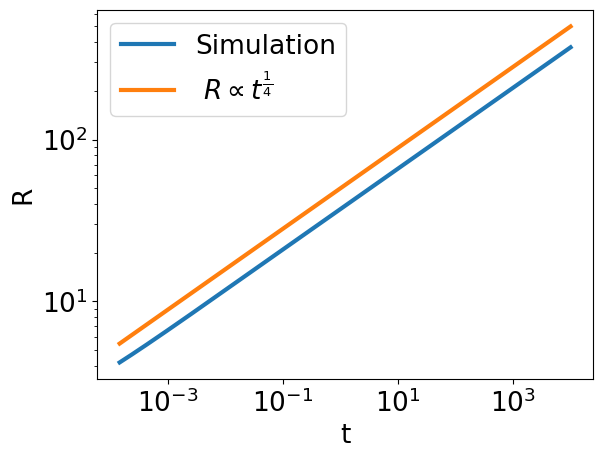}
\caption{\label{fig:radius as function of time}Radius as a function of time for $U(r)=\frac{1}{r^2}$}.
\end{figure}

Since the focus of this thesis is the dynamics of the suspension, Fig.~\ref{fig:radius as function of time} plots the radius of the suspension as a function of time in a log-log plot.
Note that the radius versus time is a straight line in a log-log plot, i.e. the radius grows as a power law of time. We tested different exponents to find the best fit for the simulations. In the case above ($U(r)=\frac{1}{r^2}$), the radius grows as $R\propto t^{\frac{1}{4}}$, as shown in Fig.~\ref{fig:radius as function of time}.

\begin{figure}[H]
\centering
\includegraphics[width=0.5\linewidth]{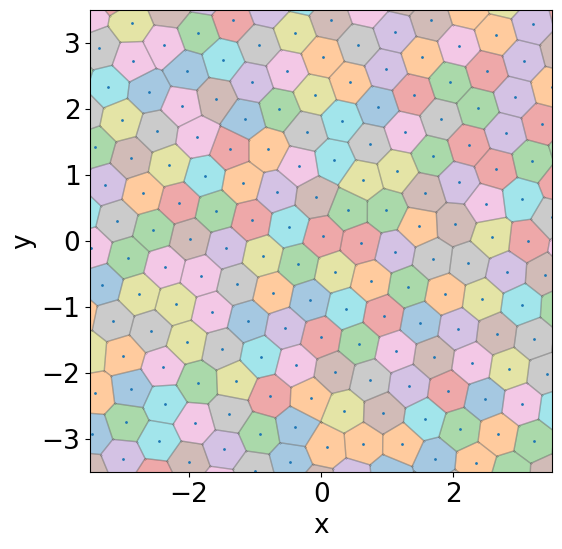}
\caption{\label{fig:voronoi diagram} Zoomed in Voronoi diagram for $N=3000$ particles, the small blue dots are the particles and the surrounding colored areas are their Voronoi area, the colors themselves are used only as a visualization aid.}
\end{figure}

Next we wanted to examine how the density profile of the suspension changes with time and distance from the center. Before presenting the results, we will first explain how the density is defined and calculated. We started by calculating the Voronoi area of each particle. The Voronoi area of a particle is defined as all the points in space that are closest to that particle than to any other particle. A visualization of the Voronoi areas of some particles in the suspension can be seen in Fig.~\ref{fig:voronoi diagram}. Next, we took the inverse of the Voronoi area of each particle and defined it as that particle's density, that is $\rho_{particle}=\frac{1}{A_{voronoi}}$. Since particles at the edge of the suspension have essentially infinite Voronoi areas, as they are not bounded by other particles beyond the edge of the suspension, we defined the density of the particles at the edge of the suspension to be zero. In Fig.~\ref{fig:voronoi diagram}, the formation of a triangular lattice structure emerging is shown. This 
triangular crystallization was first predicted by Eugene Wigner in 1934 \cite{PhysRev.46.1002}. Wigner lattices have since been proven to exist \cite{cite-key}. Wigner first stated that a low-density electron gas would form a triangular lattice since the potential energy between particles decreases slower than their kinetic energy. It is therefore more energetically favorable for particles to be located at equally distanced lattice sites in order to minimize their interaction energy. A Wigner lattice emerges in our repulsive system as well, as a result of the overdamped dynamics of the system. The potential energy behaves like $U(r) \propto \frac{1}{r^{k}}$, and the velocity is proportional to the force which is simply the gradient of the potential, such that $\mathbf{v(r)} \propto \frac{1}{r^{k+1}}$, so that in this system the condition that the potential energy decreases more slowly than the kinetic energy is also true.
Since our system is unbounded, unlike a regular Wigner lattice, the distance between lattice sites varies and defects emerge.

\begin{figure}[H]
\centering
\includegraphics[width=0.7\linewidth]{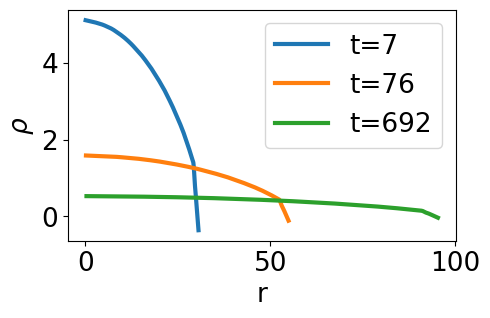}
\caption{\label{fig:density vs r} Density profile of the suspension at different time steps --- $t=7$ in blue, $t=76$ in orange and $t=692$ in green.}
\end{figure}

\begin{figure}[H]
\centering
\includegraphics[width=1\linewidth]{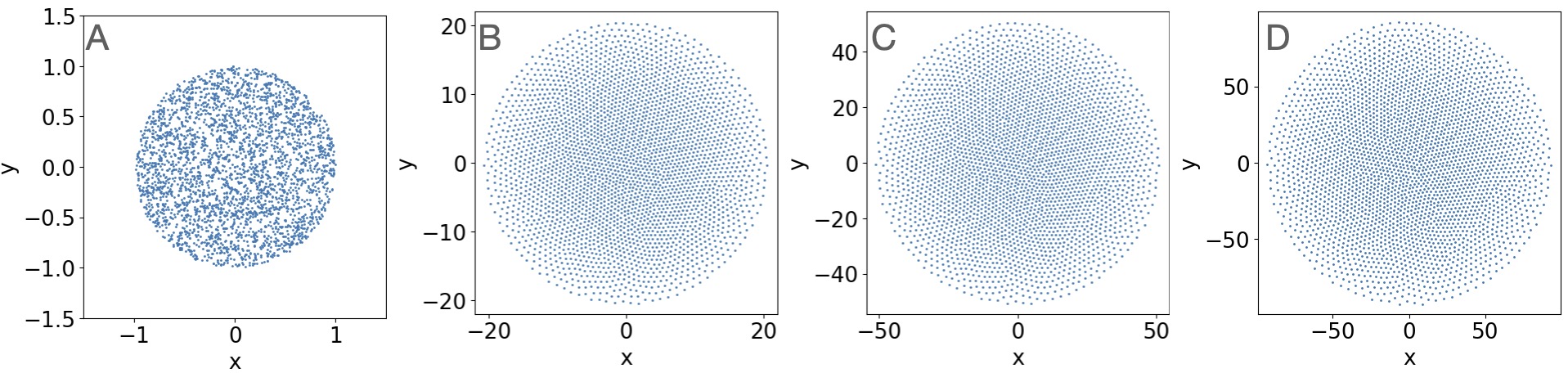}
\caption{\label{fig:initial with adaptive axis} Snapshots from Fig.~\ref{fig:inti} where the axis scales with the radius of the suspension.}
\end{figure}

Now that we have defined the particle density, we are able to plot the density profile of the suspension at different times, as shown in Fig.~\ref{fig:density vs r}
Fig.~\ref{fig:density vs r} shows the density profile of the suspension at different times. It is evident that the density decreases with time, as expected, since the particles drift further and further away from each other as time goes by. However, by re-plotting snapshots of the suspension at different times but rescaling the radius as seen in Fig.~\ref{fig:initial with adaptive axis}, we see that after some transient time, the configuration of the system does not seem to change. This suggests that the density indeed decreases with time, but the shape of the density profile remains the same. This hints that the system is self-similar.  If the system indeed exhibits self-similarity, then under the kind of re-normalization suggested in Eq.~\ref{self similar ansatz}, all the plots in Fig.~\ref{fig:density vs r} will collapse to the same curve.
To discover what the required renormalization is, we derive the coarse-grained equations for the density in the next section. 

\subsection{Governing equations}
\label{Governing equations}
In this section, we will derive the governing equation of the suspension; the derivation will be done for a general dimension $D$. But first, to get intuition to the results that follow, we will start by solving the simple case of two particles in one dimension. The equations of motion (EOM) of two particles in an overdamped limit, interacting by a power law potential $U(r)\propto \frac{1}{r^{k}}$ can be written as,
\begin{align}
\label{eom for 2 particles}
&v_{1}=\frac{dx_{1}}{dt}=\mu F = \mu \frac{1}{|x_{1}-x_{2}|^{k+1}}\\
&v_{2}=\frac{dx_{2}}{dt}=-\mu F = -\mu \frac{1}{|x_{1}-x_{2}|^{k+1}}. \notag
\end{align}
Subtracting one equation from the other yields,
\begin{align}
\label{solving eom for 2 particles}
&v_{1}-v_{2}=\frac{d(x_{1}-x_{2})}{dt}=2\mu F = 2\mu \frac{1}{|x_{1}-x_{2}|^{k+1}}\\
&\frac{dr}{dt}=2\mu \frac{1}{|r|^{k+1}}\notag\\
&r\propto t^{\frac{1}{k+2}},
\end{align}
where we defined $r=x_{1}-x_{2}$. Eq.~\ref{solving eom for 2 particles} shows that the distance between two particles interacting via a power law potential grows with time as $t^{\frac{1}{k+2}}$, we advise the reader to remember this value of $\frac{1}{k+2}$, as it will surface again many times in this work.
We will now examine the behavior of a suspension of many particles, interacting by a power law potential $U(r)\propto \frac{1}{r^{k}}$ in an overdamped system.
The density of the suspension of particles is governed by the continuity equation, which in general dimension $D$ is
\begin{equation}
\label{con eq for sim}
\    \frac{\partial \rho}{\partial t} + \frac{1}{r^{D-1}}\frac{\partial (\rho r^{D-1}v)}{\partial r}=0,
\end{equation}
where $\rho$ is the density of particles and $v$ is the radial velocity.
In order to prove that a system is self-similar, it is sufficient to show that the density function can be expressed in the form $\rho(r,t)=At^{\gamma}f\left(\frac{Br}{t^{\beta}}\right)$. To find the exponents $\gamma$ and $\beta$, the self-similar ansatz, Eq.~\ref{self similar ansatz}, is plugged into Eq.~\ref{con eq for sim} along with Eq.~\ref{discrete velocity}. It is important to remember that the constraint given by Eq.~\ref{Constraint_N} is also applicable in this system, as the number of particles in the simulations does not change.

The value of $\gamma$ was previously found in Eq.~\ref{finding gamma} to be $\gamma=-\beta D$, which is still valid here. Now, we will find $\beta$ for a general potential $U(r)\propto \frac{1}{r^{k}}$. We will begin by taking Eq.~\ref{discrete velocity} to the continuum limit, but first we will define the density as $\rho(r) = \sum_{j}^{N} \delta(\boldsymbol{r_j} - \boldsymbol{r})$
\begin{align}
    \label{continouum velocity}
       & \mathbf{v_{i}(r_{i})}=\mu\mathit{\sum_{j\neq i}^{N}}\frac{1}{\left|\mathbf{r_{i}-r_{j}} \right|^{k+1}}\mathbf{\hat{r}_{ij}}\notag\\ => &\mathbf{v(r)}=\mu\int_{0}^{R}\frac{\mathbf{r-r'}}{\left|\mathbf{r-r'} \right|^{k+2}}\rho(\boldsymbol{r'})d^{D}r'.
\end{align}
We will now make the following transformation from $(r,t)$ to $\eta$ where $\eta=\frac{Br}{t^{\beta}}$.
\begin{align}
    \label{continouum velocity eta depenednce}
    \mathbf{v(r)}&=\mu \int_{0}^{R}\frac{\mathbf{r-r'}}{\left|\mathbf{r-r'} \right|^{k+2}}\rho(\boldsymbol{r'})r'^{D-1}\boldsymbol{dr'}\notag\\
    &= \mu \int_{0}^{\frac{BR}{t^{\beta}}}\frac{\boldsymbol{\eta-\eta'}}{\left|\boldsymbol{\eta-\eta'} \right|^{k+2}}At^{-\beta D}f\left(\boldsymbol{\eta'}\right)\frac{\frac{t^{\beta(D+1)}}{B^{D+1}}}{\frac{t^{\beta(k+2)}}{B^{k+2}}}\eta'^{D-1}\boldsymbol{d\eta'}\notag\\
    &= \mu At^{-\beta(k+1)}B^{k-D+1}\int_{0}^{\eta_{B}}\frac{\boldsymbol{\eta-\eta'}}{\left|\boldsymbol{\eta-\eta'} \right|^{k+2}}f(\boldsymbol{\eta'})\eta'^{D-1}\boldsymbol{d\eta'}.
\end{align}
Where the upper boundary of integration is now $\eta_{B}=\frac{BR}{t^{\beta}}$.
We will now plug the self-similarity ansatz in Eq.~\ref{con eq for sim}  as well. We will start with the LHS,
\begin{align}
\label{1st term}
    \frac{\partial \rho}{\partial t} = -\beta At^{-\beta D -1}\left( D f(\eta) + \eta \frac{d f}{d \eta}\right).
\end{align}
The RHS transforms as
\begin{align}
\label{2nd term}
   & \frac{1}{r^{D-1}}\frac{\partial (r^{D-1}nv)}{\partial r}=  \frac{ \mu  B^{k
   -D+2}A^{2}t^{-\beta(D+k+2)}}{\eta^{D-1}}\frac{d(f(\eta)\eta^{D-1} I)}{d \eta} ,
\end{align}
where we defined 
\begin{align}
    \label{integration part of velocity}
    \boldsymbol{I} = \int_{0}^{\eta_{B}}\frac{\boldsymbol{\eta-\eta'}}{\left|\boldsymbol{\eta-\eta'} \right|^{k+2}}f(\eta')\eta'^{D-1}\boldsymbol{d\eta'}.
\end{align}
Since we transformed to the parameter $\eta$ both terms must now be independent of $(r,t)$ however in Eq.~\ref{1st term} and \ref{2nd term} there is still some dependence on $t$, so to ensure there is no $t$ dependence, we will choose $\beta$ such that the $t$ dependence 
vanishes after the transformation. Equating the $t$ dependence of both terms Eq.~\ref{1st term} and \ref{2nd term} yields,
\begin{align}
\label{value of beta in thesis}
    t^{-\beta(D+k+2)}\propto t^{-\beta D-1}& =>  -\beta(D+k+2)=-\beta D-1\\
                      & \beta = \frac{1}{k+2}.\notag
\end{align}
Notice that even when working with a general dimension $D$, the $\beta$ derived is independent of $D$. Thus the scaling of the density with $\beta$ is always given by Eq.~\ref{value of beta in thesis}, independent of the dimension. Also note that, as was said in Section \ref{Riesz Gas}, $k$ can take values from $(-2,\infty]$ since a negative $\beta$ is not physical as it implies that $r\rightarrow\infty$ is equivalent to $t\rightarrow\infty$ and not $t\rightarrow 0$. In Ref.~\cite{lewin2022coulomb}, there is a different explanation for this constraint, coming from the fact that for the long-ranged interaction, the potential needs to be a positive type. A positive type means that it has a positive Fourier transform. The Fourier transform of $1/r^{k} $ as written in \cite{lewin2022coulomb} is given by
\begin{align}
    \mathcal{F}(\frac{1}{r^{k}})=\frac{2^{\frac{D}{2}-k}\Gamma \left(\frac{D-k}{2}\right)}{\Gamma \left(\frac{k}{2}\right)} \mathcal{H}\left(\frac{1}{s^{D-k}}\right),
\end{align}
where $D$ is the dimension, $\Gamma$ is the gamma function and $\mathcal{H}\left(\frac{1}{s^{D-k}}\right)$ is the Hadamard finite part \cite{hadamad}.
The denominator $\Gamma (\frac{k}{2})$ changes sign at every non-positive even integer, which leads to the constraint that $k>-2$. 
It is quite interesting that the same constraint on $k$ is achieved from vastly different considerations.

\begin{figure}[H]
\centering
\includegraphics[width=0.8\linewidth]{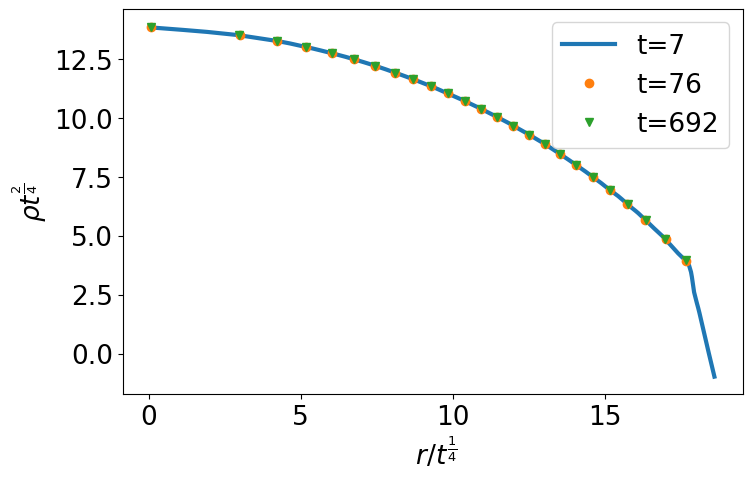}
\caption{\label{fig:density vs eta} Density profile of the suspension after renormalization according to Eq.~\ref{self similar ansatz}}
\end{figure}

\begin{figure}[H]
\centering
\includegraphics[width=0.8\linewidth]{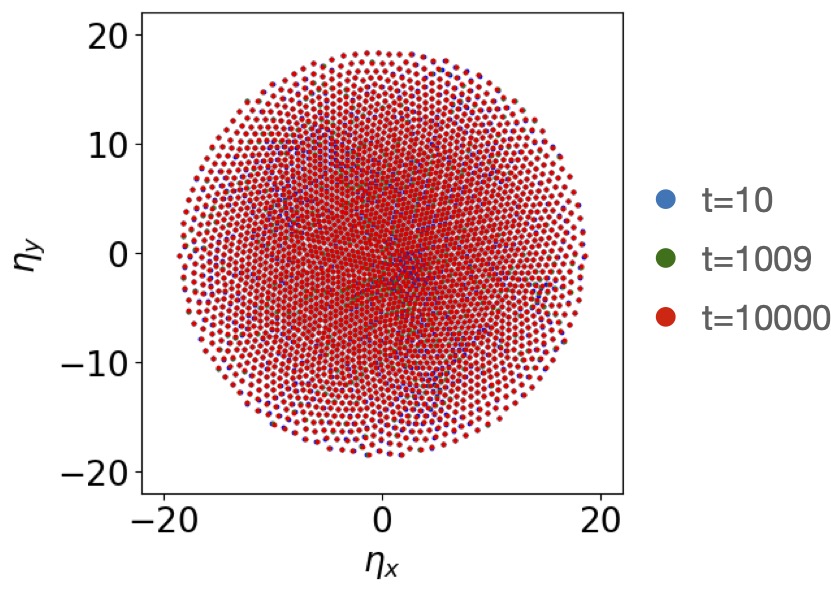}
\caption{\label{fig:normalized spreading} Renormalized spreading of particles according to $\eta=\frac{Br}{t^{\beta}}$ at three different times.}
\end{figure}

Now that we have found that $\beta=\frac{1}{k+2}$ (Eq.~\ref{value of beta in thesis}) and $\gamma=-\beta D$ (Eq.~\ref{finding gamma}), the density of the suspension can be renormalized according to the self-similar ansatz presented in Eq.~\ref{self similar ansatz}. In Fig.~\ref{fig:density vs eta}, we verify that the theory is correct and the density profile is self-similar since all plots collapse to a single curve when rescaling correctly. Fig.~\ref{fig:normalized spreading} shows the 
renormalized spreading of the suspension at three different times, where the axis are now in $\eta = \frac{Br}{t^\beta}$ coordinates. We can notice here as well that under said renormalization, all particles mostly remain at the same place, further illustrating that the system has self-similar properties.  
Plugging $\beta$ (Eq.~\ref{value of beta in thesis}) and $\gamma$ (Eq.~\ref{finding gamma}) along with the self-similarity ansatz (Eq.~\ref{self similar ansatz}), into the continuity equation (Eq.~\ref{con eq for sim}) , the following governing equation for $f(\eta)$ is achieved,
\begin{align}
\label{governing eq for f in thesis before find I}
    &-\frac{d}{d\eta}\left[\eta^{D}f(\eta)\right] + \mu B^{k-D+2}A(k+2)\left[\frac{d}{d\eta}\left(\eta^{D-1}f(\eta)I\right)\right]=0 \\
    &\eta = \alpha \left| \int_{0}^{\eta_{B}} \frac{\eta - \eta'}{\left|\eta - \eta'\right|^{k+2}} f(\eta') \eta'^{D-1} d\eta' \right|, \notag
\end{align}
where we have integrated once and the integration constant must be zero in order for $f(\eta)=0$ to be a valid solution, and we defined $\alpha=\mu B^{k-D+2}A(k+2)$. Eq.~\ref{governing eq for f in thesis before find I} can be solved numerically or analytically under certain assumptions. In the next section we will discuss the analytic solution. Note that even without the solution for the distribution we can say two things exactly --- In a suspension of repulsive particles with a power-law potential $U \sim 1/r^k$: 
\begin{itemize}
    \item  the radius of the suspension grows with time as $t^{1/(k+2)}$, independent of the dimension. 
    \item The density profile is self- similar with $\eta \propto r/t^{1/(k+2)}$ and its amplitude decreases with time as $t^{-D/(k+2)}$.
\end{itemize}
In Sec.~\ref{different density profiles}, we will verify these two results from the simulations, and in Sec.~\ref{experimental results}, we will show preliminary results verifying the first point in an experiment. 

\subsection{Finding $f(\eta)$ for $k>D$}
In this section, we will solve Eq.~\ref{governing eq for f in thesis before find I} analytically for short-ranged interactions. This will result in finding an expression for $f(\eta)$. The first step in solving Eq.~\ref{governing eq for f in thesis before find I}, is to solve Eq.~\ref{integration part of velocity}, which can be challenging since it depends on $f(\eta)$ in itself. We will begin by writing Eq.~\ref{integration part of velocity} again,
\begin{align}
    \label{integration part of velocity eta dependence}
    \boldsymbol{I}= \int_{0}^{\eta_{B}}\frac{\boldsymbol{\eta-\eta'}}{\left|\boldsymbol{\eta-\eta'} \right|^{k+2}}f(\boldsymbol{\eta'})d^{D}\eta'.
\end{align}

\begin{figure}[H]
\centering
\includegraphics[width=0.9\linewidth]{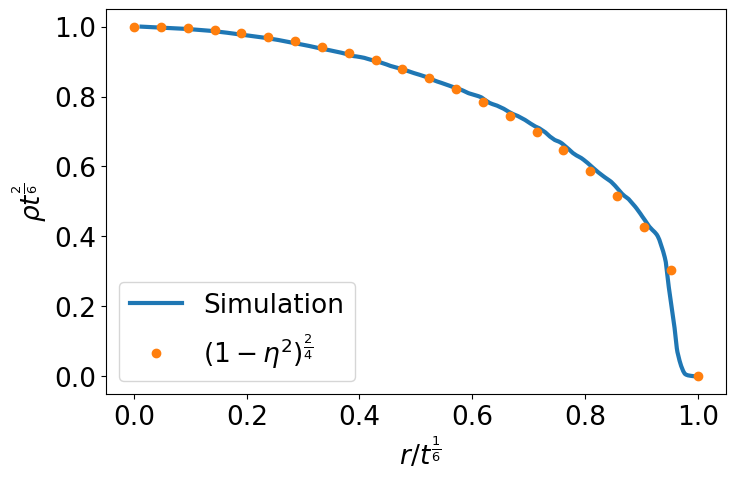}
\caption{\label{fig:moriera vs simulation k=4} Density profile from simulation with $k=4$ ($\beta=1/6$), alongside analytical solution Eq.~\ref{f in thesis}. The data is normalized by the maximum value in each axis.}
\end{figure}

To solve Eq.~\ref{integration part of velocity eta dependence} we will follow the same considerations made in Ref.~\cite{PhysRevLett.132.238201}. By assuming short-range interaction ($k>D$) and the fact that there are no particles beyond the edge of the suspension, we can extend the integration boundaries in Eq.~\ref{integration part of velocity eta dependence} to the entire space. Then by preforming a multipole expansion of $f(\eta)$ (remembering that $f(\eta)$ represents the density),
\begin{align}
    \label{multipole expansion of f}
    f(\boldsymbol{\eta'})=f(\boldsymbol{\eta}+\boldsymbol{s})\approx f(\boldsymbol{\eta})+\boldsymbol{s}\cdot \vec{\nabla} f(\boldsymbol{\eta}),
\end{align}
where $\boldsymbol{s}=\boldsymbol{\eta-\eta'}$. 
Inserting Eq.~\ref{multipole expansion of f} into Eq.~\ref{integration part of velocity eta dependence} while also extending the integration bounds to the entire space results in,
\begin{align}
\label{velocity in terms of s}
    \boldsymbol{I} = -S_{D-1}\vec{\nabla} f(\boldsymbol{\eta}) \cdot \int_{0}^{\infty}\frac{\hat{s}\hat{s}}{s^{k+1}}s^{D}ds,
\end{align}
where $S_{D-1}$ is the surface area of a $D-1$ dimensional unit sphere. The first moment in Eq.~\ref{multipole expansion of f} vanishes after integration from symmetry considerations. In order to solve Eq.~\ref{velocity in terms of s} it is first crucial to resolve the divergence from the lower boundary. Moriera \textit{et. al.} \cite{PhysRevE.98.032138}~\cite{PhysRevE.93.060103} , helps us in this regard. Moriera explains that it is highly unlikely for two repulsive particles in a dissipative medium to collide, so this divergence is not physical. To resolve this divergence, an excluded area around each particle can be defined, which increases/decreases with the decrease/increase in density. This excluded area can be approximated by the density, $r_{exc}=\alpha f^{-\frac{1}{D}}$. By incorporating this excluded area, Eq.~\ref{velocity in terms of s} can be written as,
\begin{align}
\label{velocity in terms of s with moriera}
    \boldsymbol{I} = -S_{D-1}\vec{\nabla} f(\boldsymbol{\eta}) \int_{\alpha f^{-\frac{1}{D}}}^{\infty}s^{D-k-1}ds=S_{D-1}\vec{\nabla} f(\boldsymbol{\eta})\frac{\alpha^{D-k}}{D-k}f^{\frac{k}{D}-1}.
\end{align}
Using Eq.~\ref{velocity in terms of s with moriera} we can write Eq.~\ref{2nd term} as
\begin{align}
    \label{2nd term with moriera}
    \frac{1}{r^{D-1}}\frac{\partial (nr^{D-1}v)}{\partial r}=- \frac{S_{D-1} \mu B^{k-D+2}A^{2}\alpha^{D-k}}{\eta^{D-1}(k-D) }\frac{d}{d \eta}(f(\eta)^{\frac{k}{D}}\eta^{D-1}  \frac{d f}{d \eta}) ,
\end{align}
where we have also used $\beta=\frac{1}{k+2}$ from Eq.~\ref{value of beta in thesis} to remove the $t$ dependence. Now plugging Eq.~\ref{2nd term with moriera} and Eq.~\ref{1st term} into Eq.~\ref{con eq for sim} and setting $B^{k-D+2}=\frac{ (k-D)}{(k+2)S_{D-1} \mu A \alpha^{D-k}}\frac{k}{2D}$, the following governing equation for $f(\eta)$ is achieved,
\begin{align}
\label{governing eq for f in thesis}
     \frac{d}{d\eta}\left[\eta^{D}f(\eta)\right] + \frac{k}{2D}\left[\frac{d }{d \eta}\left(\eta^{D-1} f^{\frac{k}{D}}(\eta)\frac{d f}{d \eta}\right)\right]=0.
\end{align}
We can notice that Eq.\ref{governing eq for f in thesis} is exactly the same equation that was solved in Sec.~\ref{Non Linear Diffusion} Eq.~\ref{governing eq for f in non linear diff} with $n=\frac{k}{D}$. The solution to Eq.~\ref{governing eq for f in thesis} is given by Eq.~\ref{f for non linear diff},
\begin{align}
    \label{f in thesis}
    f(\eta)=(1-\eta^{2})^{\frac{D}{k}}.
\end{align}
This solution for $f(\eta)$ given by Eq.~\ref{f in thesis} has what is known as "compact-support" — the solution is strictly zero beyond some cutoff.
Plotting Eq.~\ref{f in thesis} alongside the density profile for the case of $k=4$ such that the condition $k>D$ is satisfied. Fig.~\ref{fig:moriera vs simulation k=4} shows an excellent agreement between the simulation and the analytical solution derived here. We can also observe that the two plots drift a bit from each other at the tails. This may be the result of coarse-graining done in the simulation, namely that some particles at the edge of the suspension have identically zero particle density.

It is important to remember for the next section that the solution for $f(\eta)$ seen in Eq.~\ref{f in thesis} is only valid for values of $k>D$ since otherwise the integral in Eq.~\ref{velocity in terms of s} does not converge.

\subsection{Density profiles for different values of $k$}
\label{different density profiles}
So far, we have shown simulations for a system of particles interacting with each other with a potential $U(r) \propto \frac{1}{r^{2}}$. Such a system was observed to have self-similar properties, satisfying the self-similar ansatz specified in Eq.~\ref{self similar ansatz}, and particles were centered at the origin. Now, we will investigate how different values of $k$ lead to significantly different density profiles.

The behavior of the system when $k=2$ was already discussed in previous sections, so it will be revisited only briefly here. The focus now will be on three other cases in two dimensions --- first, for $k=4$ (i.e. $k>D-2$), in which case we will get an origin-centered density. Secondly, For $k=0$, i.e., where $U(r)\propto \log(r)$, we will show that this is the limiting case ($k = D-2$) and the density is constant. Lastly, for $k=-1$ (i.e. $k<D-2$), where we will show particles are centered at the boundary. Systems that interact via a log potential are often referred to as "log gas". In $2D$ they are also known as a "Coulomb gas" \cite{Minnhagen1987TheTC}. 
In general, we observe that the density profile is: 
\begin{itemize}
    \item origin-centered for $k>D-2$.
    \item constant profile at $k=D-2$.
    \item boundary-centered for $k<D-2$.
\end{itemize}

Note that these values of $k$ are different from those that characterize short-ranged ($k> D$) and long-ranged ($k \leq D$) behaviors. The short versus long-ranged nature is related to the behavior of the potential energy and whether it is finite or diverges at infinity, whereas the density has a stricter condition. A long-ranged potential with $k=2$ in $2D$ is centered at the origin (as we have shown in Fig.~\ref{fig:density vs eta}), as well as for $k=1$. Whereas its long-ranged pears $k=0$ and $k=-1$ behave differently as we show below. Additionally, we will see that for $k=4$, an origin-centered density also emerges, which means that both a long-ranged potential ($k=(2,1)$), and a short-ranged potential ($k=4$), result in an origin-centered density. This leads us to the conclusion that the short/long range classification is not sufficient to distinguish the different density profiles that emerge. We remind the readers that values of $k\leq-2$ are not physical, as we have shown in Sec.~\ref{Governing equations}. 
\begin{figure}
\centering
\includegraphics[width=0.8\linewidth]{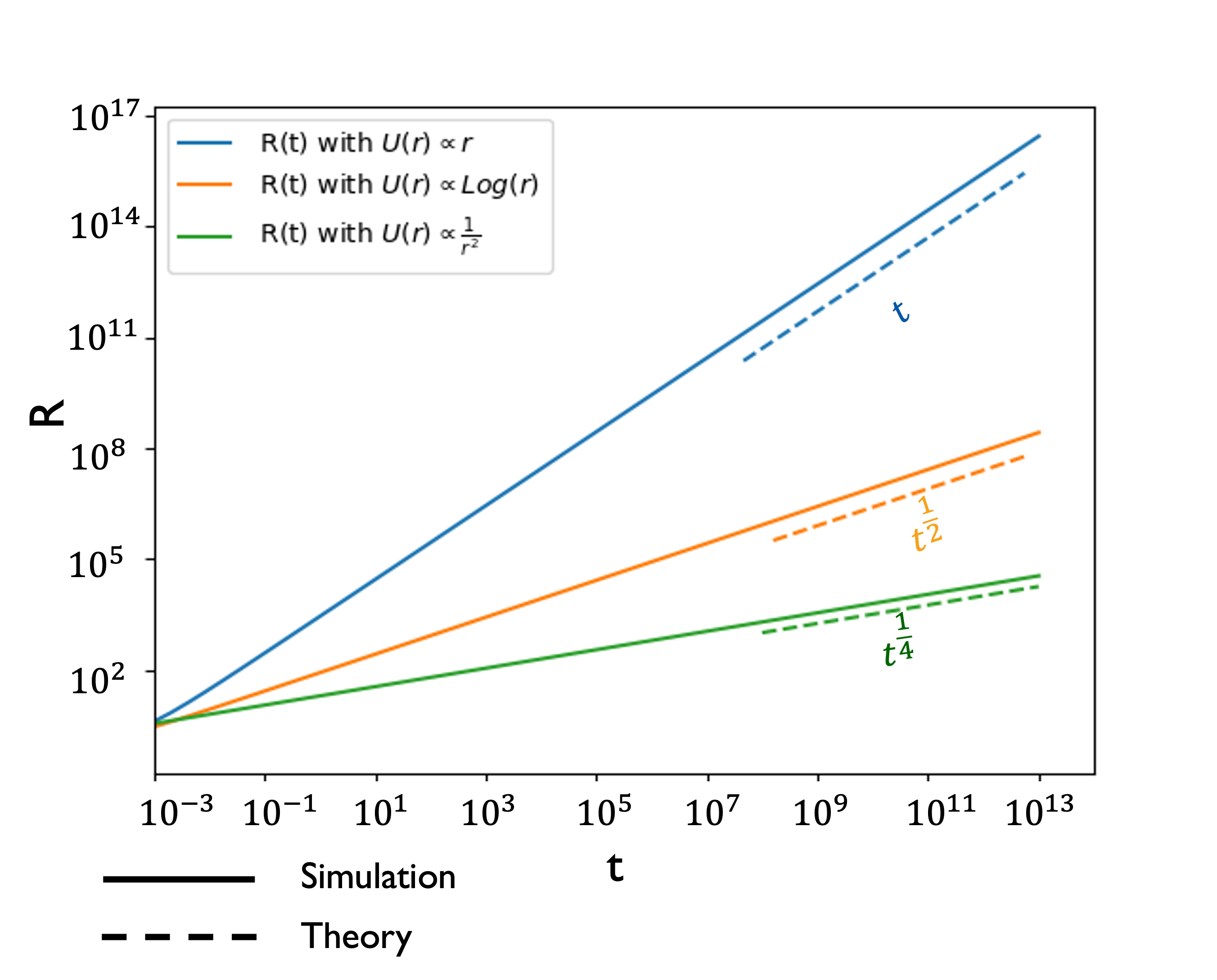}
\caption{\label{fig:growth for different values of k} $R(t) \propto t^{\frac{1}{k+2}}$ for different values of $k$ (-1,0,2).}
\end{figure}
First, we will examine how the suspension behaves for different values of $k$. From Eq.~\ref{value of beta in thesis}, we know that the radius should grow as $t^{\frac{1}{k+2}}$, irrespective of the dimension. In Fig.~\ref{fig:growth for different values of k}, it is shown that for all three values of $k$ specified before, the radius follows the expected power of $t$ as predicted by the theory derived in this thesis. This is insightful because it allows one to understand the growth of an overdamped system of particles interacting through a power law potential $U(r) \propto \frac{1}{r^{k}}$ by simply knowing the value of $k$. We can also notice in Fig.~\ref{fig:growth for different values of k} that for $k=0$ i.e. $U(r) \propto \log(r)$, the suspension grows with the same time dependence as the square root of the mean squared displacement (MSD) of a diffusive drop, namely $\sqrt{MSD}\propto t^{\frac{1}{2}}$, even though for pure diffusion there are no interactions between particles, and here the particles are purely repulsive. We can extend this analogy to more values of $k$, some of which can be seen in Fig.~\ref{fig:growth for different values of k}.
\begin{itemize}
    \item subdiffusion $k>0$.
    \item regular diffusion for $k=0$.
    \item superdiffusive for $-1<k<0$
    \item ballistic for $k=-1$.
\end{itemize}

\begin{figure}
\centering
\includegraphics[width=0.85\linewidth]{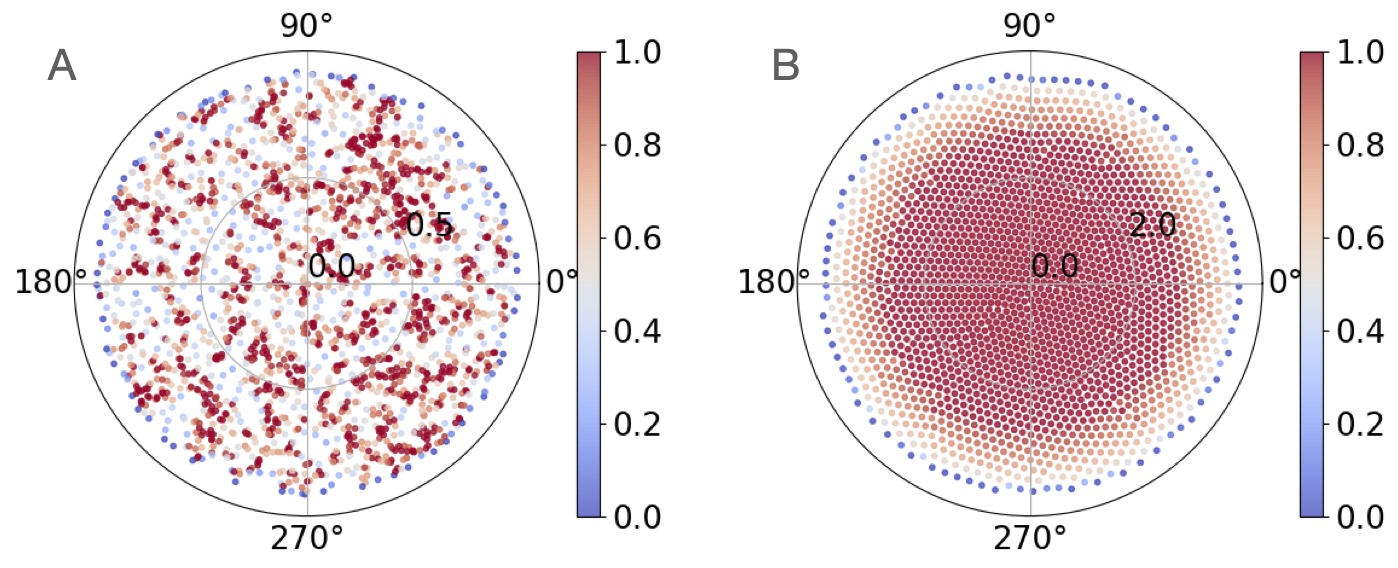} 
\caption{\label{fig:heatmap for k=4} Heat map for the density of the suspension for $k=4$. The scale bar is normalized by the mean of the densities, and the $r$ axis grows with the suspension. A) initial configuration. B) 
 after 1000 timesteps.}
\end{figure}

\begin{figure}
\centering
\includegraphics[width=0.7\linewidth]{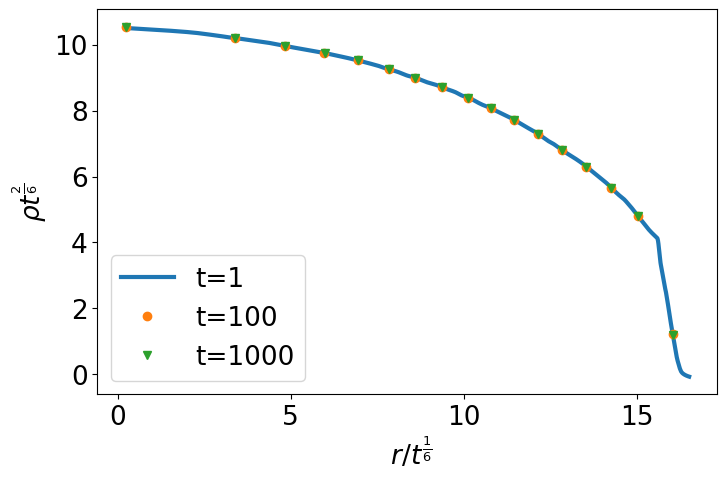} 
\caption{\label{fig:normalized density for k=4} Re-normalized density according to self simillar ansatz Eq.~\ref{self similar ansatz} for $k=4$ i.e. $\beta = \frac{1}{6}$, $\gamma =-\frac{2}{6}$}
\end{figure}

\begin{figure}[H]
\centering
\includegraphics[width=0.85\linewidth]{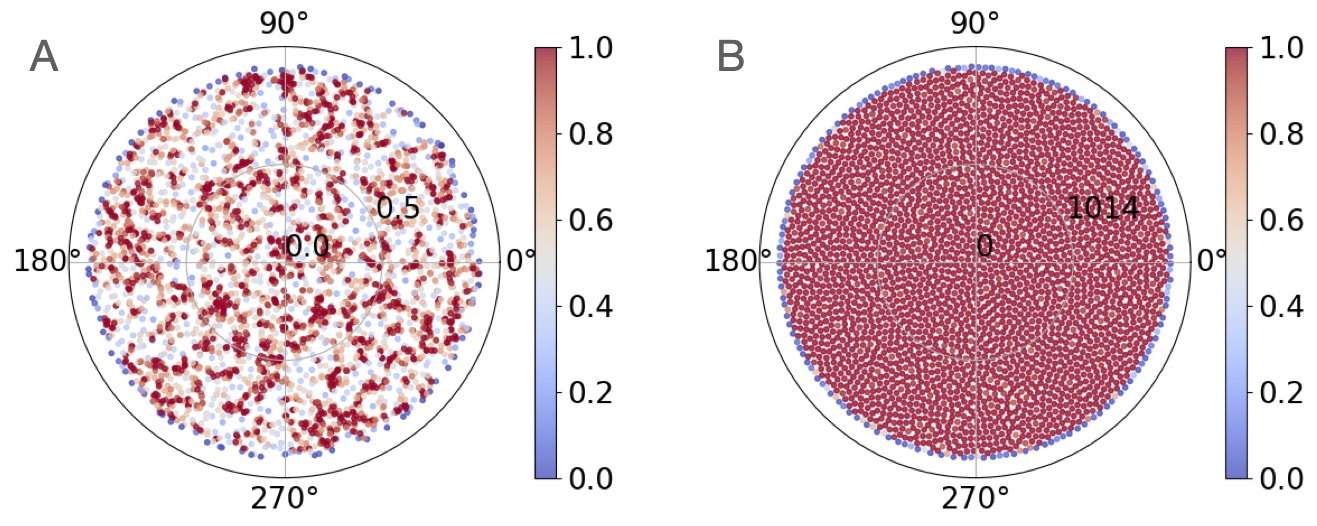} 
\caption{\label{fig:heatmap for k=0} Heat map for the density of the suspension for $k=0$. The scale bar is normalized by the mean of the densities, and the $r$ axis grows with the suspension. A) initial configuration. B) 
 after 100 timesteps.}
\end{figure}

\begin{figure}[H]
\centering
\includegraphics[width=0.7\linewidth]{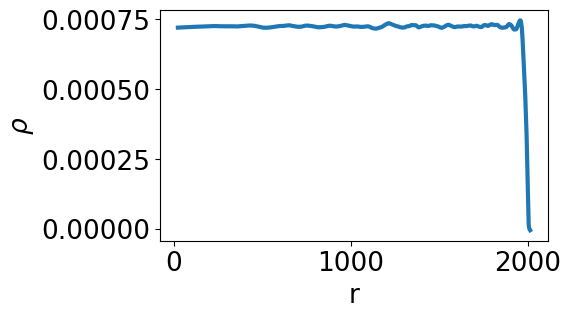} 
\caption{\label{fig:density for k=0} Density profile for $k=0$ after 100 time steps.}
\end{figure}

Next, the density profile for the potentials where $k=(4,0,-1)$ will be analyzed, as well as their self-similar properties. The heat map of the suspension can be seen in Fig.~\ref{fig:heatmap for k=4}, which shows an origin-centered density as well as a Wigner lattice Formation. The density profile was shown back in Fig.~\ref{fig:moriera vs simulation k=4}, whereas the self-similar properties can be seen in Fig.~\ref{fig:normalized density for k=4}.\\
Fig.~\ref{fig:heatmap for k=0} shows that the resulting density of the suspension for the case of $k=0$, remains constant, as further illustrated in Fig.~\ref{fig:density for k=0}. This is quite different from the density profile shown in Fig.~\ref{fig:density vs r}

\begin{figure}[H]
\centering
\includegraphics[width=0.85\linewidth]{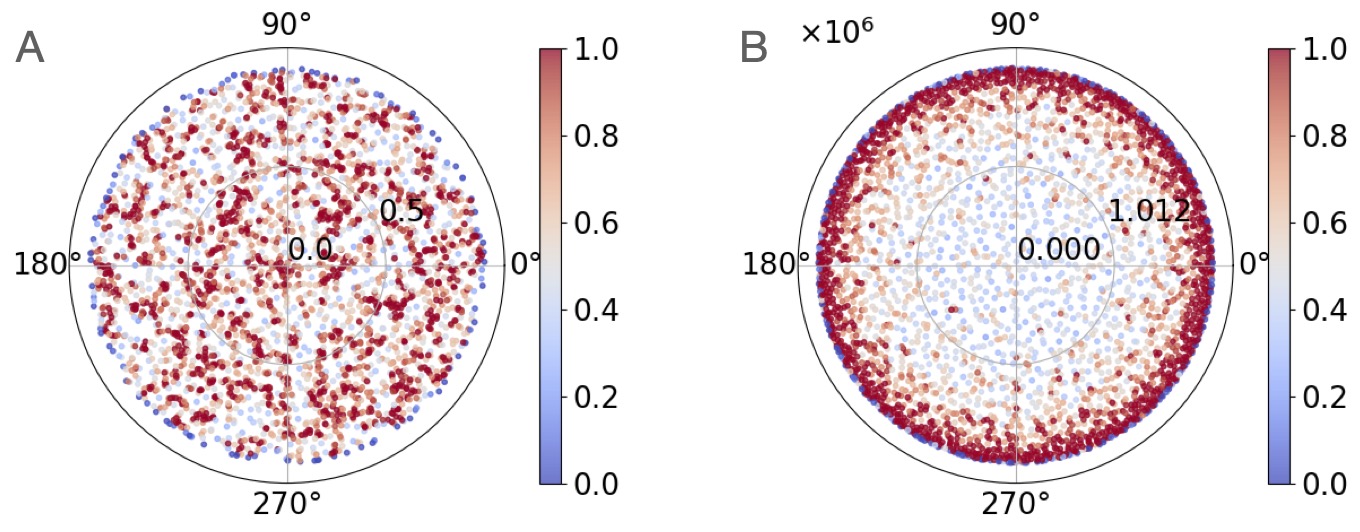} 
\caption{\label{fig:heatmap for k=-1} Heat map for the density of the suspension for $k=-1$, the scale bar is normalized by the mean of the densities and the $r$ axis grows with the suspension. A) initial configuration. B) after 100 timesteps.}
\end{figure}

\begin{figure}[H]
\centering
\includegraphics[width=0.7\linewidth]{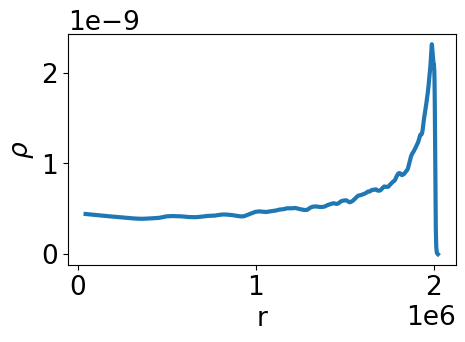} 
\caption{\label{fig:density for k=-1} Density profile for $k=-1$ after 100 time steps.}
\end{figure}

Now, taking $k=-1$ i.e $U(r) \propto r $. This indicates that the force between particles is constant in magnitude and does not depend on the distance between them, but only on the direction of the force. In Fig.~\ref{fig:heatmap for k=-1}, a distinct density profile is observed compared to the previous cases $k=(4,2,0)$. For $k=-1$,  the particles tend to accumulate at the boundary of the suspension. This behavior is also illustrated in Fig.~\ref{fig:density for k=-1}, where a sharp peak in density is evident at the edge of the suspension.

\begin{figure}[H]
\centering
\includegraphics[width=0.7\linewidth]{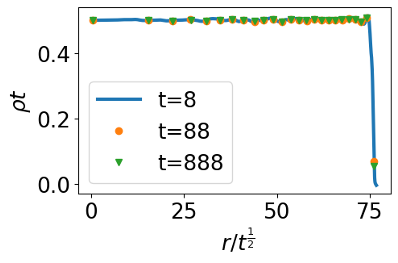} 
\caption{\label{fig:normalized density for k=0} Re-normalized density according to self simillar ansatz Eq.~\ref{self similar ansatz} for $k=0$ i.e. $\beta = \frac{1}{2}$, $\gamma =-1$}
\end{figure}

\begin{figure}[H]
\centering
\includegraphics[width=0.7\linewidth]{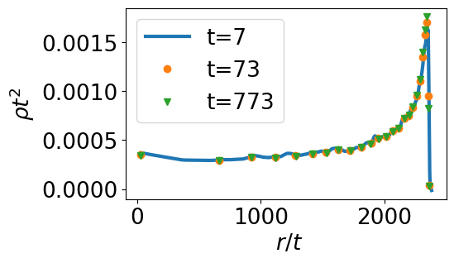} 
\caption{\label{fig:normelized density for k=-1} Re-normalized density according to self simillar ansatz Eq.~\ref{self similar ansatz} for $k=-1$ i.e $\beta = 1$, $\gamma =-2$}
\end{figure}

Figures~\ref{fig:normalized density for k=0} and~\ref{fig:normelized density for k=-1} verify that the density is self-similar, as it is evident that the density satisfies the self-similar ansatz specified in Eq.~\ref{self similar ansatz}. The density curves at different time steps fall on a single curve under the relevant re-normalization. 
Notice the vastly different density profiles resulting from using different values of $k$. There is an origin centered density profile when $k=2$ (long-ranged) Fig.~\ref{fig:density vs r} or when $k=4$ (short-ranged) Fig.~\ref{fig:normalized density for k=4}, a constant density profile for $k=0$ (long-ranged) Fig.~\ref{fig:density for k=0} and a boundary centered density profile for when $k=-1$(long-ranged) Fig.~\ref{fig:density for k=-1}.

It is clear that the short/long-ranged classification is insufficient to categorize the emerging density profiles. We aimed to categorize which values of $k$ correspond to their profiles. By simulating different values of $k$, we discovered that what determines the shape the density profile of the suspension is whether $k$ is smaller,larger or equal to $D-2$ where $D$ is the dimension. We find that for $k>D-2$, an origin centered profile emerges, for $k=D-2$, a constant density profile appears, and for $k<D-2$, a boundary-centered profile appears. Notice that with this new classification, both a short-ranged potential ($k=4$) and a long-ranged potential ($k=(2,1)$), exhibit the same density profiles for $D=2$. Moreover, three vastly different density profiles can emerge for long-ranged potentials Fig.~\ref{fig:density vs r},Fig.~\ref{fig:density for k=0} and Fig.~\ref{fig:density for k=-1}. 
Let us note that when $k=D-2$, the potential is known as the Coulomb potential. Additionally, the boundary-centered density profile is referred to in the literature as the "Evaporation Catastrophe" \cite{gallavotti1999statistical} and is attributed to the potential being too repulsive at infinity.

\subsection{Finding $f(\eta)$ for $k\leq D$}
Let us try to give some intuition to the unique profiles 
in Fig.~\ref{fig:normalized density for k=0} and Fig.~\ref{fig:normelized density for k=-1}.  
As was already pointed out, Fig.~\ref{fig:normalized density for k=0} indicates that for $k=D-2$, $f(\eta)$ is constant. We will use this knowledge to show that a constant profile solves Eq.\ref{governing eq for f in thesis before find I} for $k=D-2$.
We start from the equation for the density, 
\begin{align}
\label{finding f for k=D-2}
    &\eta=\alpha \left| \int_{0}^{\eta_{B}}\frac{\boldsymbol{\eta-\eta'}}{\left|\boldsymbol{\eta-\eta'} \right|^{k+2}}f(\eta')\boldsymbol{d\eta'}\right|.
\end{align}
Let us define $\boldsymbol{s}=\boldsymbol{\eta'}-\boldsymbol{\eta}$. Taking $f={\rm const}$, we can take it out of the integral, such that 
\begin{align}
    & \frac{\eta}{f}=\alpha \left| \int_{-\eta}^{\eta_{B}-\eta} \frac{\boldsymbol{s}}{s^{k+2}}d^Ds\right| = \alpha S_{D-1} \left| \int_{-\eta}^{\eta_{B}-\eta} \frac{{\rm sign}(s)}{s^{k+1}}s^{D-1}ds\right|  \notag \\
    & \frac{\eta}{\alpha S_{D-1} f}= \left| \int_{-\eta}^{0} {\rm sign}(s)s^{D-2-k}ds + \int_{0}^{\eta_{B}-\eta} {\rm sign}(s)s^{D-2-k}ds\right| \notag \\
    & \frac{\eta}{\alpha S_{D-1} f}= \left| \int_{-\eta}^{0} s^{D-2-k}ds - \int_{0}^{\eta_{B}-\eta} s^{D-2-k}ds\right| \notag \\
    & \frac{\eta}{\alpha S_{D-1} f}= \left| \int_{-\eta}^{0} ds - \int_{0}^{\eta_{B}-\eta} ds\right| = 2\eta - \eta_{B} \notag\\
    & \frac{d}{d\eta}\left(\frac{\eta}{\alpha S_{D-1} f}\right) = \frac{d}{d\eta}(2\eta - \eta_{B}) \notag \\
    & f=\frac{1}{2\alpha S_{D-1}}. \notag
\end{align} 
We got that indeed $f={\rm const}$ is a solution, showing that our assumption is valid for $k=D-2$.
We have yet to solve Eq.\ref{governing eq for f in thesis before find I} for other values in the regime of $k<D$, however, we have found a candidate solution for this regime in 1D given in \cite{PhysRevLett.123.100603}. They show that for such a power-law potential in a harmonic trap, the solution has the following form,
\begin{align}
    \label{f for k<D in 1D}
    f(\eta)\propto (1-\eta^{2})^{\frac{k+1}{2}}.
\end{align}
Notice how Eq.~\ref{f for k<D in 1D} has the same form as Eq.~\ref{f in thesis} with the difference being the power law that governs the function $(1-\eta^{2})$, i.e. in Eq.~\ref{f in thesis} the power is $\frac{D}{k}$ while in Eq.~\ref{f for k<D in 1D} the power is $\frac{k+1}{2}$.

\begin{figure}[H]
\centering
\includegraphics[width=0.8\linewidth]{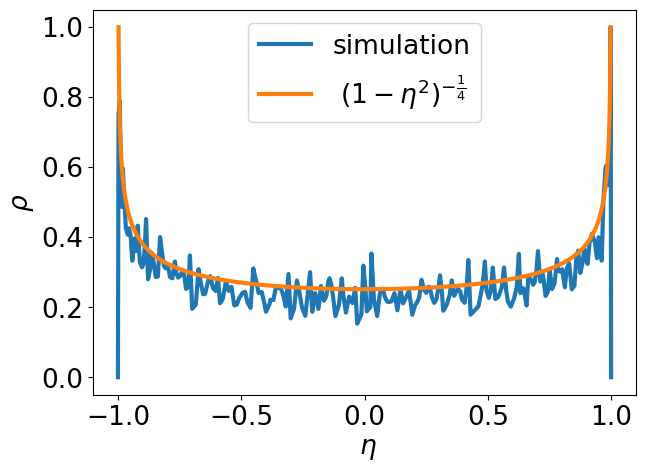} 
\caption{\label{fig:f for k<D in 1D} $f(\eta)$ from  for $k=-1.5$ in $1D$.}
\end{figure}

Fig.~\ref{fig:f for k<D in 1D} shows excellent agreement between Eq.~\ref{f for k<D in 1D} and simulations. There are two things that are worth noting in the derivation of Eq.~\ref{f for k<D in 1D} \cite{PhysRevLett.123.100603}. The first being that the system which was studied in \cite{PhysRevLett.123.100603} indeed has particles interacting with each other with a power law potential, however, they are also subjected to an external harmonic potential, unlike in this work where the particles are spreading in free space with no external potential to confine them. Additionally, Eq.~\ref{f for k<D in 1D} was derived using a statistical mechanics approach, which requires the system to be in equilibrium. In contrast, our work uses the continuity equation and focuses on the dynamics of the system which is inherently out of equilibrium. We are currently working on generalizing Eq.~\ref{f for k<D in 1D} to higher dimensions.

\subsection{Collision between suspensions}

\begin{figure}[H]
\centering
\includegraphics[width=1\linewidth]{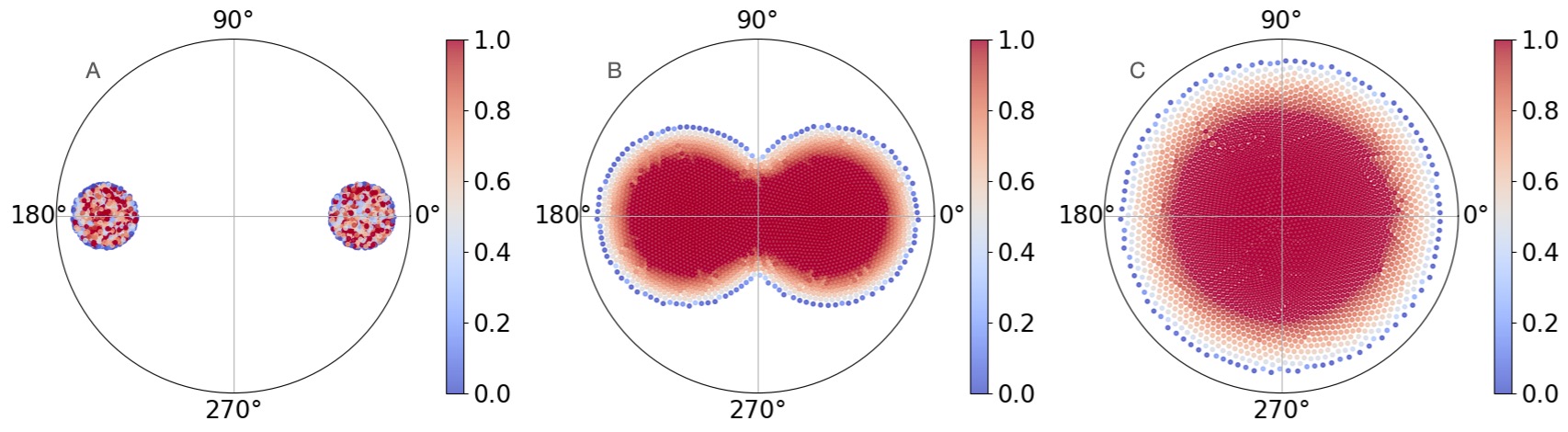} 
\caption{\label{fig:heat map for 2 suspensions k=3} Collision of two suspensions of $N=1500$ particles each for $k=3$. The image shows the two suspensions: A) At the initial configuration.B) Colliding and starting to merge, C) Merging into one isotropic suspension.}
\end{figure}

So far, we have only looked at the case of a single suspension expanding in free space. Now, we will examine 
qualitatively what happens when two or more suspensions collide. We will simulate the collision of two suspensions by initiating the simulation with two or more suspensions separated by a certain distance, so that they do not touch each other initially. l then evolve the simulation in time. At some point, the suspensions meet. We ask, what is the density of the suspension as a function of time. For an absorbing equation, the initial conditions do not matter, and the state at long times is always the same. We will show that when $k\geq D-2$, the equation is absorbing, whereas for $k<D-2$, the system remembers its initial state. 
We will begin with the case where $k=3$. As we have shown, single circular suspension, results in an origin-centered isotropic density profile.
In Fig.~\ref{fig:heat map for 2 suspensions k=3}, we observe that when two suspensions interact with a potential of the form $U(r)\propto \frac{1}{r^{3}}$, they begin to merge upon collision, ultimately resulting in a single suspension without any discernible indication of the original two suspensions. Additionally, we notice that following the merger, the density profile remains centered at the origin as expected from section \ref{different density profiles}. 

\begin{figure}[H]
\centering
\includegraphics[width=1\linewidth]{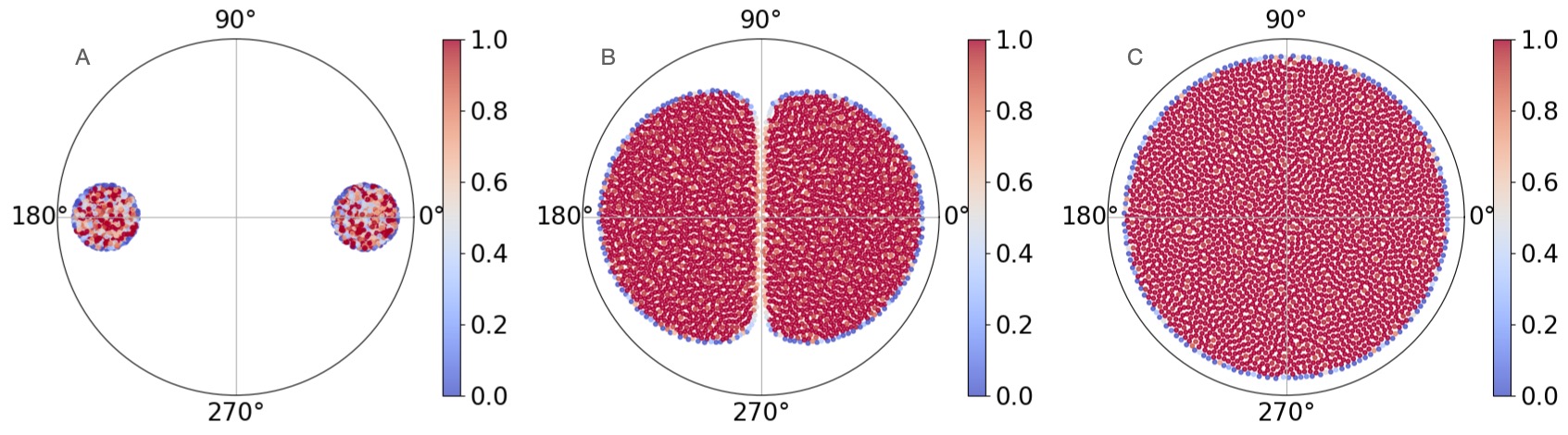} 
\caption{\label{fig:heat map for 2 suspensions k=0} Collision of two suspensions of $N=1500$ particles each for $k=0$. The image shows the two suspensions:A) At the initial configuration.B) Colliding and small repulsion between the suspension.  C) Merging into one suspension.}
\end{figure}

Next, we will examine the case of $k=0$, where $U(r)\propto \log{(r)}$. A single suspension had a uniform density profile. 
Comparing the collision process when $k=3$ and $k=0$, we observed that in Fig.~\ref{fig:heat map for 2 suspensions k=3}, the merging of the two suspensions appeared seamless. However, in Fig.~\ref{fig:heat map for 2 suspensions k=0}, we noticed that the two suspensions initially repelled each other, leading to the formation of a particle-free boundary. Eventually, the two suspensions merged into one circularly symmetric suspension with a constant density profile, similar to the result in Fig.~\ref{fig:heatmap for k=0}. Upon observing the final image, it is impossible to determine whether we started with one or two suspensions. Therefore, these two potentials result in absorbing final states. 

Lastly, we will look at the case of $k=-1$ that is $U(r)\propto r$.
In this case, we observed a significant difference from the previous two cases (see Fig.\ref{fig:heat map for 2 suspensions k=3} \ref{fig:heat map for 2 suspensions k=0}).Most notably, the two suspensions did not merge. A pronounced particle-free zone is maintained upon collision and always remains. In the case of $k=0$, a boundary had formed in Fig.~\ref{fig:heat map for 2 suspensions k=0}, but it quickly disappeared. 
\begin{figure}[H]
\centering
\includegraphics[width=\linewidth]{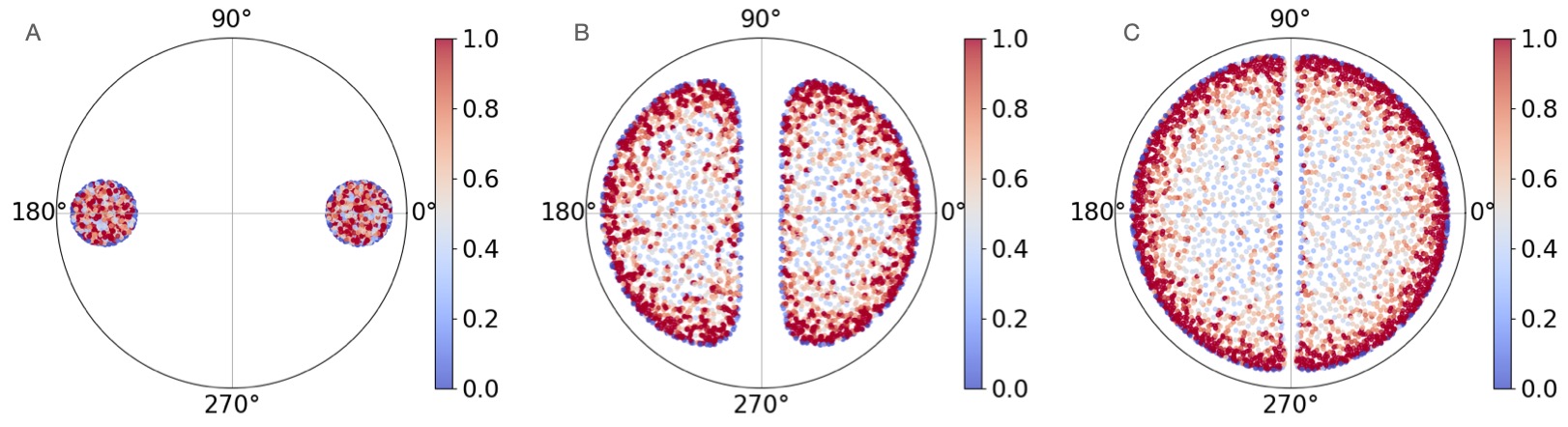} 
\caption{\label{fig:heat map for 2 suspensions k=-1} collision of two suspensions of $N=1500$ particles each for $k=-1$.Image shows the two suspensions: A) At the initial configuration. B)  Starting to repel one another. C) Fully repel each other and not merging.}
\end{figure}

\begin{figure}[H]
\centering
\includegraphics[width=0.8\linewidth]{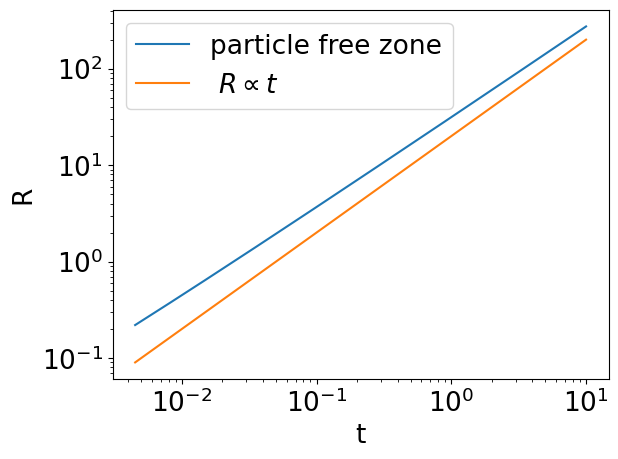} 
\caption{\label{fig:particle free zone between suspensions} Width of particle free zone shown in Fig.~\ref{fig:heat map for 2 suspensions k=-1} plot along time, $R\propto t$ is also plotted to five a reference for the slope. }
\end{figure}

\begin{figure}[H]
\centering
\includegraphics[width=0.4\linewidth]{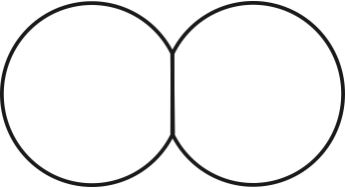} 
\caption{\label{fig:2 soap bubbles} Two soap bubbles forming a boundary where they meet.}
\end{figure}

In Fig.~\ref{fig:particle free zone between suspensions} the distance between the two suspensions in Fig.~\ref{fig:heat map for 2 suspensions k=-1} is plotted over time, this is essentially the width of the particle free zone that was shown. There are two interesting things that we can learn from Fig.~\ref{fig:particle free zone between suspensions}, first, the distance between the suspensions (and the width of the particle free zone) grows with time. second, the distance between the suspensions grows a bit more slowly with time than the radius of an individual suspension --- for $k=-1$ $\beta =1$, so $R\propto t$, and Fig.~\ref{fig:particle free zone between suspensions} shows a smaller slope. This is most likely because the particles at the boundary of the particle free zone, experience an outward force from their own suspension, while also experiencing a force from the other suspension which pushes them back into their own suspension.
Note the resemblance between Fig.~\ref{fig:heat map for 2 suspensions k=-1} and the collision of two soap bubbles Fig.~\ref{fig:2 soap bubbles}, which also forms a boundary between the bubbles. However, in soap bubbles, the interaction is attractive, leading to the aggregation of particles at the boundary. Whereas in Fig.~\ref{fig:heat map for 2 suspensions k=0}, the interaction is repulsive, and the boundary between the suspensions is particle-free.

\begin{figure}[H]
\centering
\includegraphics[width=0.8\linewidth]{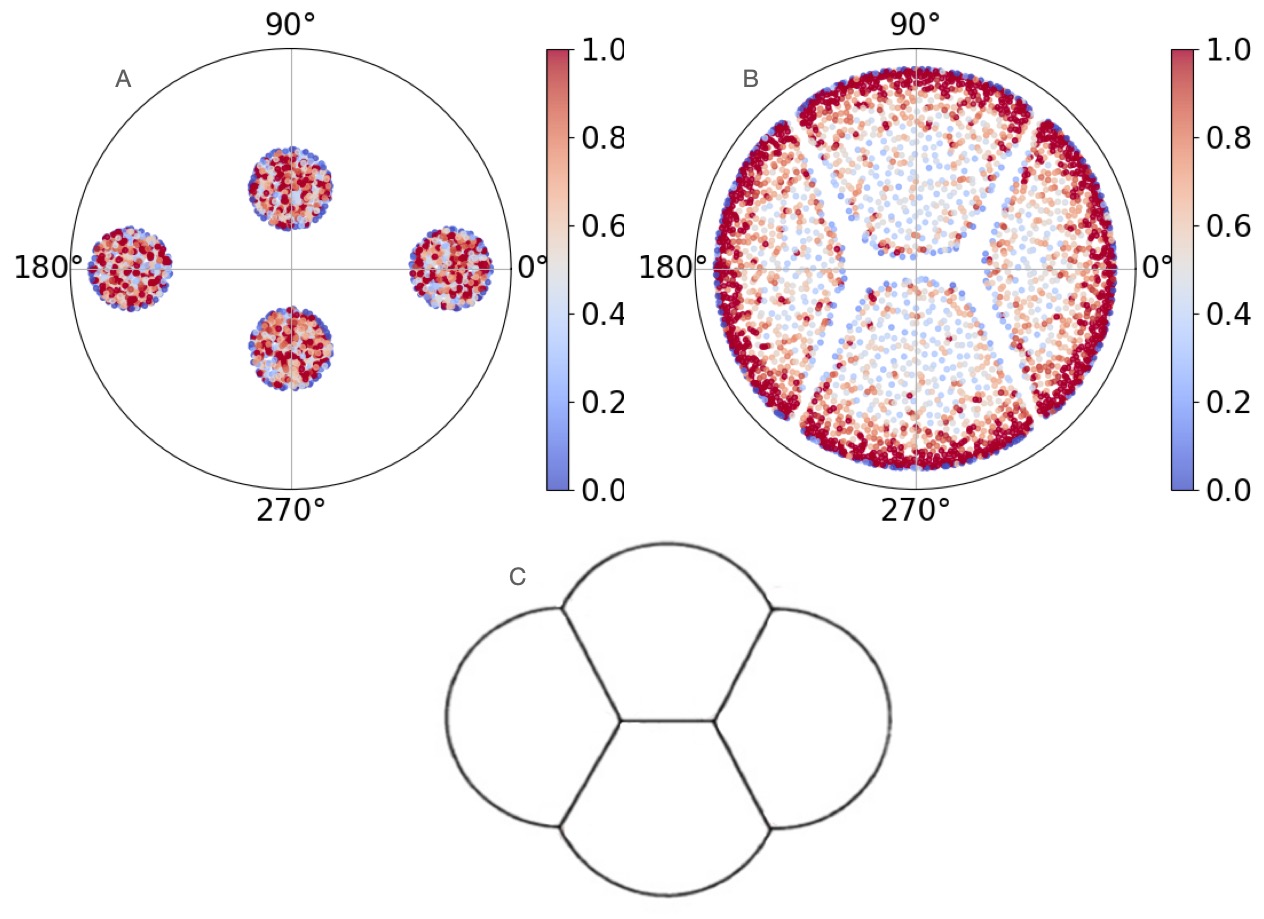} 
\caption{\label{fig:4 suspensions} A) Initial configuration of 4 suspensions of $N=750$ particles each. B) Final configuration of the 4 suspensions. C) Possible boundary configuration for 4 soap bubble.}
\end{figure}

In order to further check whether or not a similarity with soap bubbles exist, a configurations of more than two suspensions (and bubbles) were examined as well.Fig.~\ref{fig:4 suspensions} shows a collision between four suspensions, further strengthening the resemblance to soap bubbles. A more in-depth and quantitative analysis of the similarity between the two phenomena is beyond the scope of this work. Furthermore, the formation of boundaries between soap bubbles is still an open question, specifically in the case of four bubbles or more  \cite{RevModPhys.79.821}. 

\subsection{Preliminary experimental results}
\label{experimental results}
\begin{figure}[H]
\centering
\includegraphics[width=0.5\linewidth]{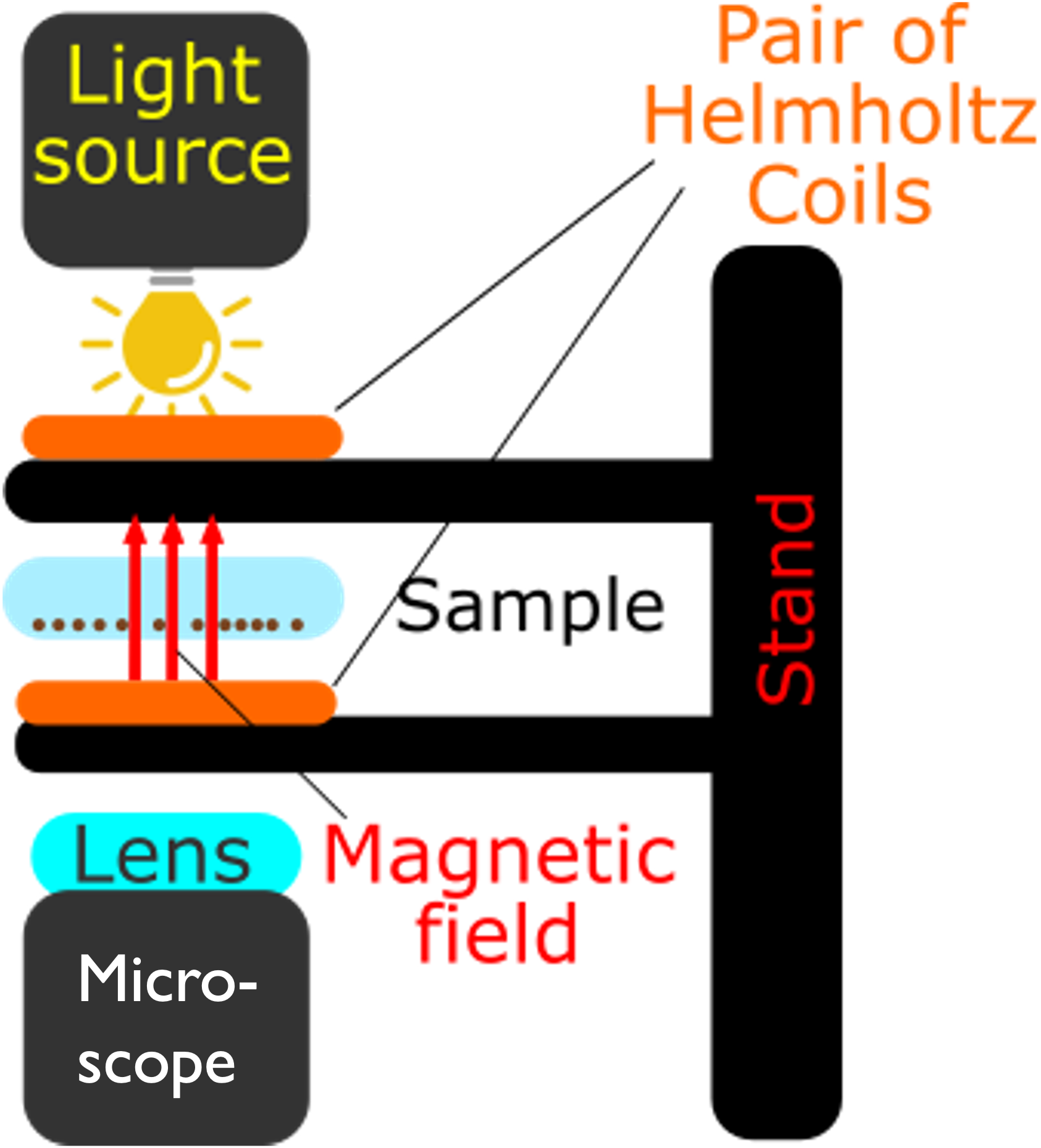} 
\caption{\label{fig:experimental setup} Illustration of the experimental setup.}
\end{figure}

\begin{figure}[H]
\centering
\includegraphics[width=1\linewidth]{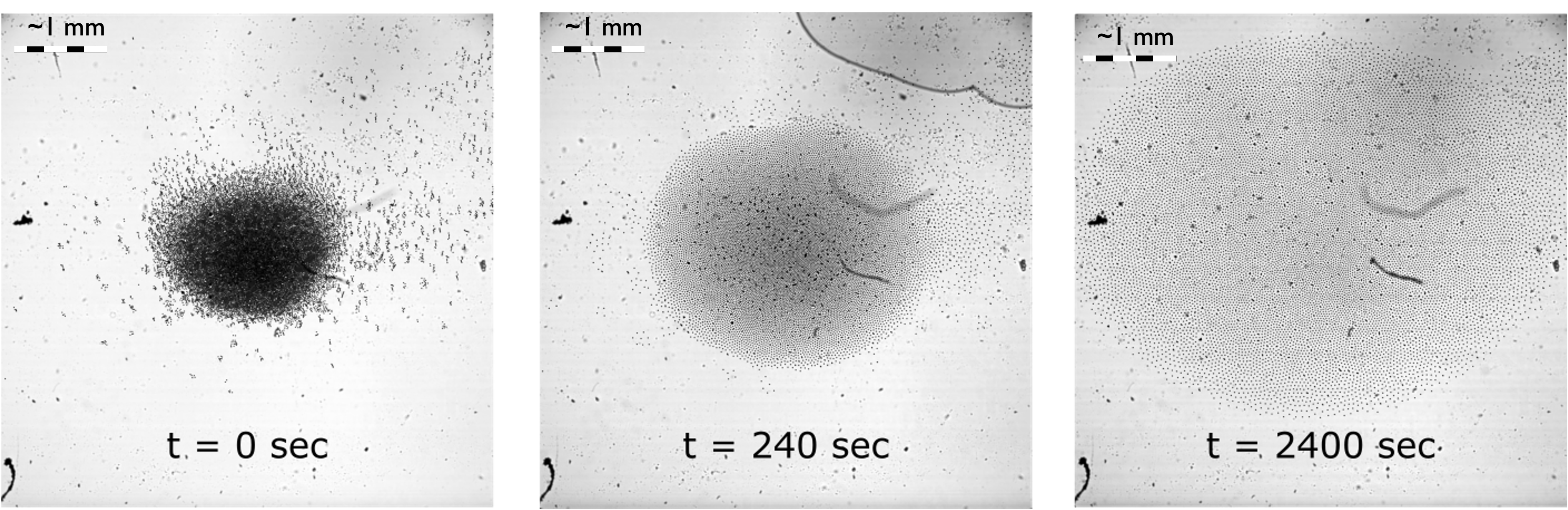} 
\caption{\label{fig:experiment timelapse} Experiment timelapse, scale bar are available at the top left corner of each photo, the time is given at the bottom center of each photo. }
\end{figure}

We are currently in the early stages of trying to validate the analytical and numerical results provided in this thesis with experiments. The experimental system we chose includes super-paramagnetic particles made of polystyrene (PS) or Silicon dioxide ($\mathrm{SiO_{2}}$) with diameters between 5-12 $\mu$m. The particles are electrostatically stabilized to avoid short-range attraction and aggregation.
We created a pair of Helmholtz coils and placed them in such a way that when current runs through them they generate a uniform magnetic field perpendicular to the plane of the sample. This, in turn, magnetizes the particles, creating dipole-dipole magnetic repulsion of the form $U(r) \propto \frac{1}{r^{3}}$. An illustration of the setup can be seen in Fig.~\ref{fig:experimental setup}.
The experiment is conducted as follows: we start by creating a sample in a capillary tube (0.2mm x 6mm x 50mm) with a ratio of 1:300 microliters of particles to deionized water, we then seal the sample using UV glue. To concentrate the particles, we use a regular magnet which has gradients in its magnetic field relative to the sample, causing the particles to congregate at the spot where the magnetic field is strongest, which we choose to be in the middle of the glass slide. Finally, we place the sample at the plane of view of the microscope (Nikon Eclipse Ti2) with 4x magnification lens and 0.2 N.A. We turn on the power supply for the Helmholtz coils, which generates a uniform magnetic field of around 4-5 mT at the plane of the sample. We then record the spreading of the particles at one frame per second, using a camera (kinetix-m-c) connected to the microscope.\\
In Fig.~\ref{fig:experiment timelapse} the time evolution of the suspension is shown. We can indeed see that the colloids repel each other, and the suspension spreads as a result. We analyzed the experiment in Python using Trackpy \cite{allan_2024_12708864} to locate the position of the colloids. We then show in Fig.~\ref{fig:growth for different values of k} the standard deviation as a function of time showing that $STD(t)\propto R(t) \propto t^{\frac{1}{k+2}}$. Since the potential is $U\propto 1/r^3$, i.e. $k=3$, the theoretical prediction is that the radius grows as $R(t)\propto t^{\frac{1}{5}}$(blue solid line in the figure), which is indeed verified. We included simulation results as a dashed orange line. 

\begin{figure}[H]
\centering
\includegraphics[width=0.8\linewidth]{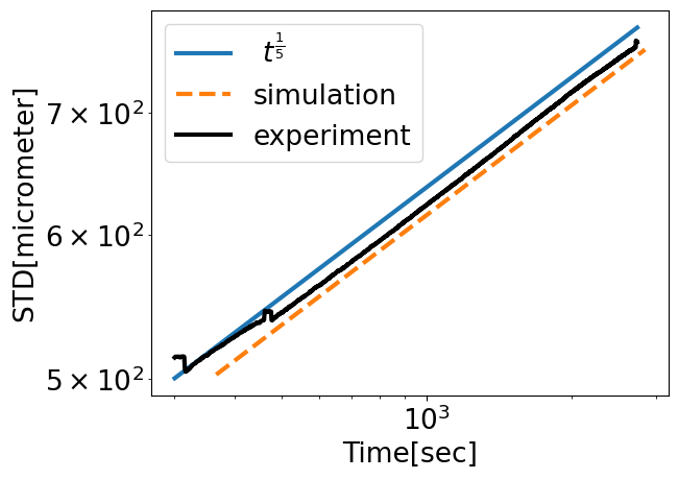} 
\caption{\label{fig:experiment growth} Standard deviation of the suspension along side numerical simulations and prediction from theory.}
\end{figure}

Fig.~\ref{fig:experiment growth} shows an excellent agreement between the simulation, theory, and experimental results. In the future, we hope to check if the experiment exhibits self-similar properties, as shown in the simulation and predicted by the theory. That is, we wish to check whether, under the relevant normalization, density plots from different times collapse to the same curve. We also plan to check colloidal particles of different sizes and compositions. 

\newpage
\section{Discussion}
In this research, we investigated an overdamped system of repulsive particles that interact through a power law potential, $U(r) \propto \frac{1}{r^k}$. We found that the radius of a suspension of such particles grows according to $R(t)\propto t^{\frac{1}{k+2}}$. We also discovered that this system exhibits self-similar properties when normalized by $\rho(r,t)=At^{\gamma}f\left(\frac{Br}{t^{\beta}}\right)$. We determined the values of $\beta$ and $\gamma$ to be $\beta=\frac{1}{k+2}$ and $\gamma =-\beta D$. With the help of Moriera et al., when a short-range potential is used ($k>D$), we found an expression for $f\left(\frac{Br}{t^{\beta}}\right)$. We observed that different density profiles arise depending on whether $k$ is greater than, less than, or equal to $D-2$. We also observed an interesting phenomenon when two or more suspensions collide for values of $k<D-2$, in which a particle-free zone is formed and the system has a memory of its initial conditions. The resulting shapes are reminiscent of soap bubbles, although further work is needed to understand the relation between the two. Finally, we have preliminary experimental results matching the theory, though this is still a work in progress.

There is still much more to investigate. In future work we hope to verify our results in this thesis with experiments, that are currently underway in our lab. We plan to study experimentally the density profile and also to experiment with different sized particles and different potentials as well.  We will also attempt to find an analytic expression for $f\left(\frac{Br}{t^{\beta}}\right)$ when a long-range potential ($k\leq D$) is used for $D>1$. In this limit, a  divergence is introduced to Eq.~\ref{velocity in terms of s} which we have yet to resolve. 
Discovering the significance of the limiting dimension $D-2$, might open a new branch of classification for repulsive potentials in dissipative media, as we have shown in Sec.~\ref{different density profiles}. 
Additionally, in this thesis we have used in our simulation point-like particles interacting by a single power law potential. In the future, we hope to simulate particles with different sizes and interacting with a potential that combines several powerlaws, which might help us understand how such systems will behave in reality.  
\newpage

\section{Supplementary}

In this supplementary, we will show results from simulations in $1D$, the graphs will be shown in the same order as their $2D$ counterparts in the main text. The initial configuration in $1D$ is that particles are uniformly distributed on a line centered around the origin.

\begin{figure}[H]
\centering
\includegraphics[width=0.7\linewidth]{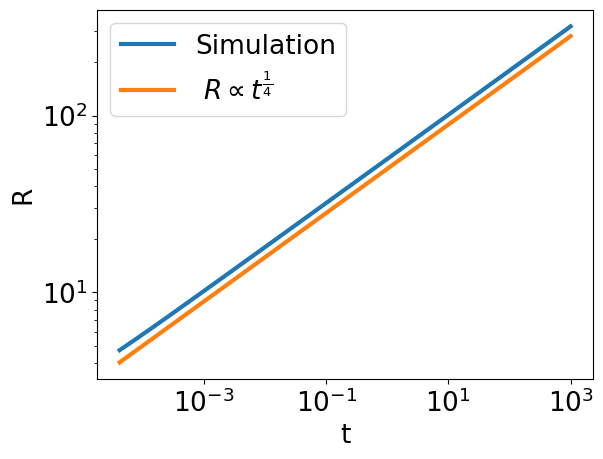}
\caption{\label{fig:R vs t d=1 k=2} $R$ as a function of time in a $1D$ suspension with $k=2$ : $\beta = \frac{1}{4}$.}
\end{figure}
Fig.~\ref{fig:R vs t d=1 k=2} shows that the size of the $1D$ suspension of particles, grows with time as $t^{\frac{1}{4}}$ for $k=2$, this is the same value that was achieved in $2D$ further showing that the growth of the suspension is independent on the dimensionality of the system.
The density in $1D$ is calculated in the same manner as in $2D$ with the exception of using the distance between particles instead of a Voronoi area --- since a $2D$ area is not defined on a $1D$ line of particles. The density of each particle is defined as half the distance from each of its neighbors.

\begin{figure}[H]
\centering
\includegraphics[width=0.7\linewidth]{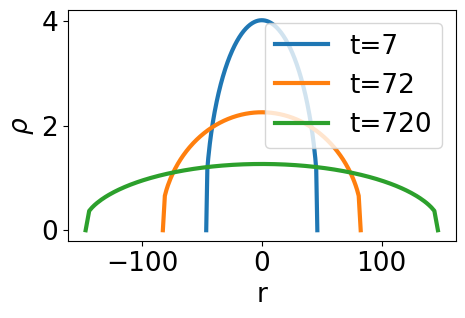}
\caption{\label{density vs r for d=1 k=2} Density profile at different times in a $1D$ suspension with k=2.}
\end{figure}

\begin{figure}[H]
\centering
\includegraphics[width=0.7\linewidth]{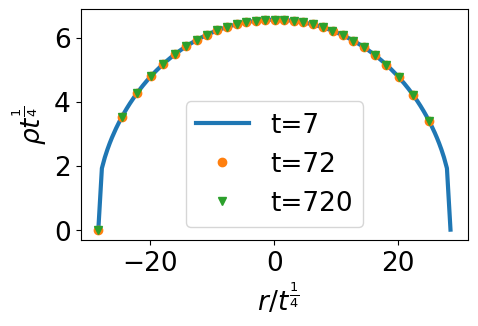}
\caption{\label{density vs eta for d=1 k=2} Re-normalized Density profile at different times in a $1D$ suspension with k=2.}
\end{figure}

\begin{figure}[H]
\centering
\includegraphics[width=0.7\linewidth]{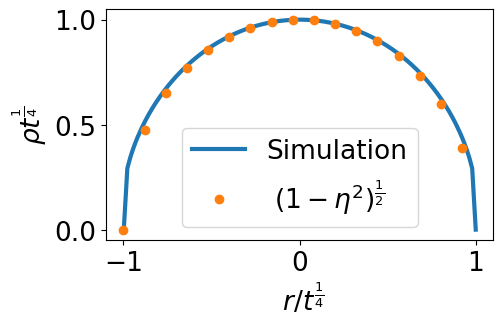}
\caption{\label{density vs eta d=1 moriera} Density profile from simulation of $1D$ suspension, with $k=2>D$  ($\beta=1/4$), alongside analytical solution. The data is normalized by the maximum value in each axis.}
\end{figure}

Fig.~\ref{density vs eta for d=1 k=2} shows that under the using the self similar ansatz $\rho(r,t)=At^{\gamma}f\left(\frac{Br}{t^{\beta}}\right)$
, all plots from Fig.~\ref{density vs r for d=1 k=2} fall to the same curve. This means that the exhibits self similar properties in $1D$ as well. This is to be expected since the derivation done in the main text is done for a general dimension $D$. In Fig.~\ref{density vs eta d=1 moriera} we can see that the solution derived for short ranged potentials ($k>D$) is satisfied in $1D$ as well.

\begin{figure}[H]
\centering
\includegraphics[width=0.7\linewidth]{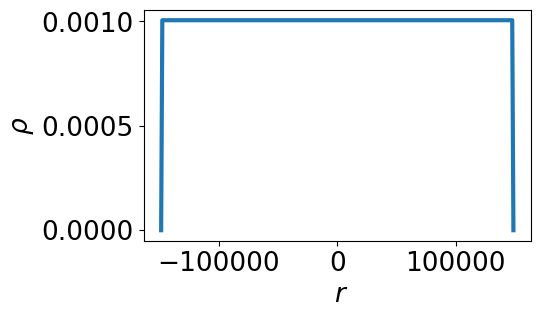}
\caption{\label{density vs r for d=1 k=-1} Density profile from simulation of $1D$ suspension, with $k=-1=D-2$ . A constant density is observed. }
\end{figure}

\begin{figure}[H]
\centering
\includegraphics[width=0.7\linewidth]{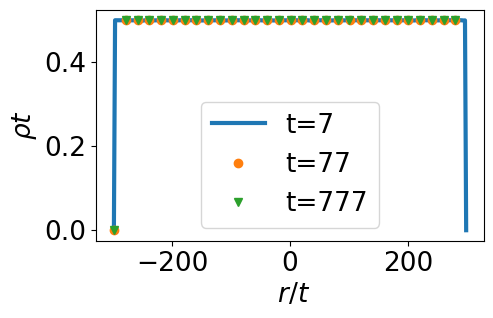}
\caption{\label{density vs eta for d=1 k=-1} Re-normalized density profile from simulation of $1D$ suspension, with $k=-1=D-2$($\beta=1$).}
\end{figure}

\begin{figure}[H]
\centering
\includegraphics[width=0.7\linewidth]{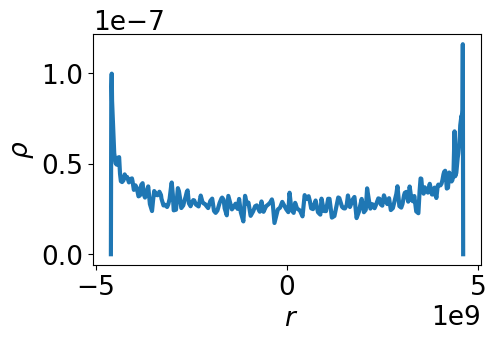}
\caption{\label{density vs r for d=1 k=-15} Density profile from simulation of $1D$ suspension, with $k=-1.5<D-2$ . A boundary centered density is observed.}
\end{figure}

\begin{figure}[H]
\centering
\includegraphics[width=0.7\linewidth]{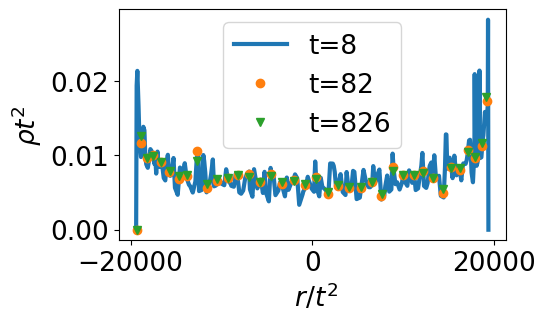}
\caption{\label{density vs eta d=1 k=-15} Re-normalized density profile from simulation of $1D$ suspension, with $k=-1.5<D-2$($\beta=2$).}
\end{figure}

In Fig.~\ref{density vs r for d=1 k=-1} and Fig.~\ref{density vs r for d=1 k=-15} we can see a constant density profile and a boundary-centered density profile in $1D$, respectively. Fig.~\ref{density vs eta for d=1 k=-1} and Fig.~\ref{density vs eta d=1 k=-15} show the self-similar properties of these profiles.

\begin{figure}[H]
\centering
\includegraphics[width=0.7\linewidth]{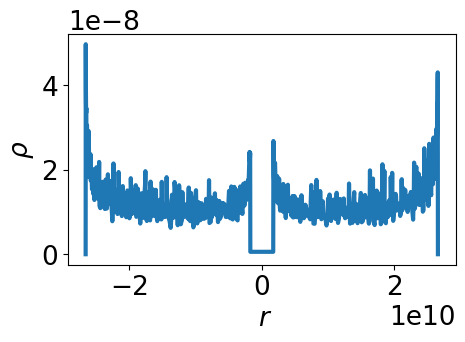}
\caption{\label{density vs r for d=1 2drops k=-15} Density profile from simulation of two $1D$ suspensions, with $k=-1.5<D-2$. A particle-free zone can be seen where the two suspensions meet.}
\end{figure}

We wanted to check if a particle free zone formed where two suspensions meet in 1D as we have observed in $2D$ for $k<D-2$. Fig.~\ref{density vs r for d=1 2drops k=-15} indeed shows that a particle free zone forms, as is evident from the density falling to zero where the suspensions meet. 

\bibliographystyle{ieeetr}
\bibliography{Main_text}

\end{document}